\documentclass[aps,prb,superscriptaddress]{revtex4}

\usepackage{graphicx}
\usepackage{epsfig}
\usepackage{amsmath}

\begin{document}

\title{On the chemical bonding effects in the Raman response:
       Benzenethiol adsorbed on silver clusters}
\author{Semion K. Saikin}
\email[]{saykin@fas.harvard.edu} \affiliation{Department of
Chemistry and Chemical Biology, Harvard University, Cambridge, MA
02138, USA.} \affiliation{Department of Physics, Kazan State
University, Kazan 420008, Russian Federation.}

\author{Roberto Olivares-Amaya}
\affiliation{Department of Chemistry and Chemical Biology, Harvard
University, Cambridge, MA 02138, USA.}

\author{Dmitrij Rappoport}
\affiliation{Department of Chemistry and Chemical Biology, Harvard
University, Cambridge, MA 02138, USA.}

\author{Michael Stopa}

\affiliation{Center for Nanoscale Systems, Harvard University,
Cambridge, MA 02138, USA.}

\author{Al\'an Aspuru-Guzik}
\email[]{aspuru@chemistry.harvard.edu} \affiliation{Department of
Chemistry and Chemical Biology, Harvard University, Cambridge, MA
02138, USA.}

\begin{abstract}
We study the effects of chemical bonding on Raman scattering from
benzenethiol chemisorbed on silver clusters using time-dependent
density functional theory (TDDFT). Raman scattering cross sections
are computed using a formalism that employs analytical derivatives
of frequency-dependent electronic polarizabilities, which treats
both off-resonant and resonant enhancement within the same scheme.
In the off-resonant regime, Raman scattering into molecular
vibrational modes is enhanced by one order of magnitude and shows
pronounced dependence on the orientation and the local symmetry of
the molecule.  Additional strong enhancement of the order of
$10^2$ arises from resonant transitions to mixed metal--molecular
electronic states.  The Raman enhancement is analyzed using Raman
excitation profiles (REPs) for the range of excitation energies
$1.6-3.0$~eV, in which isolated benzenethiol does not have
electronic transitions. The computed vibrational frequency shifts
and relative Raman scattering cross sections of the
metal--molecular complexes are in good agreement with experimental
data on surface enhanced Raman scattering (SERS) for benzenethiol
adsorbed on silver surfaces. Characterization and understanding of
these effects, associated with chemical enhancement mechanism, may
be used to improve the detection sensitivity in molecular Raman
scattering.
\end{abstract}

\maketitle
\renewcommand{\thefootnote}{\fnsymbol{footnote}}

\section{Introduction}
\label{sec:intro}

Raman scattering from molecules in proximity to a rough
noble-metal surface or near a metal nanoparticle is strongly
enhanced due to the interaction with surface plasmon modes and to
the formation of metal--molecular complexes. \cite{Moskovits_Rev,
Otto_Rev} This phenomenon, which allows for the measurement of
Raman spectra of extremely low concentrations of molecules
\cite{Meixner} with a single-molecule detection as its ultimate
limit, \cite{Nie_SM, Kneipp_SM} is very attractive for sensor
applications. In particular, it could be utilized for detection
and identification of hazardous materials
\cite{SERSdetect1,SERSdetect2} or for probing of biological
structures, \cite{SERSbio} for which Raman fingerprints provide
unique information about molecular composition.

Recent achievements in the design of nanostructured materials
demonstrate substantial progress toward resolving the
long-standing problem of low reproducibility of SERS substrates.
Arrays of nanoantennas fabricated with electron beam lithography
\cite{Ken} or nanoimprinting \cite{Chou} utilize extended control
of optically excited surface plasmons to tune the plasmon
resonance frequency and to focus the near field in particular
areas. Lithographically engineered \cite{VD_str, Reinhard} or
laser-engineered \cite{Eric} structures allow the observation of a
spatially homogeneous enhancement of the Raman signal by 7 orders
of magnitude \cite{EF_note} when an excitation laser frequency far
below intramolecular resonances is employed. However, there
remains a number of open questions which are related to the
formation of hot spots and to the homogeneity of the Raman
response throughout a sample. The enormous Raman cross section
reported for fluorescent dyes in hot spots \cite{Nie_SM} has a
substantial contribution from an intramolecular resonant
excitation. \cite{VD_sm} On the other hand, many important
analytes do not have electronic excitations in the available range
of laser frequencies. For these systems, a relevant question is
whether modification of the local environment of the molecule can
be used to boost the limit of Raman detection sensitivity.

The importance of the so-called chemical enhancement
\cite{Moskovits_Rev,ChemEnh2,ChemEnh3} has been debated since the
early years after the discovery of SERS. The contributions of the
electromagnetic coupling and the chemical binding are not clearly
separable in the experiments, \cite{ChemEnh3} because the first
process is essential for SERS detection. This adds to ambiguity in
the interpretation of SERS experiments. Moreover, the electronic
coupling effects are sensitive to the local environment of
adsorbed molecules, which makes a systematic analysis of chemical
effects more difficult. General theories of chemical enhancement
\cite{Persson, Adrian, Birke} provide an intuitive picture of the
mechanism, but rely on phenomenological parameters. In this
context, first-principles modeling \cite{Schatz_Rev} is a useful
tool that complements experimental studies and provides additional
information about microscopic properties of a metal
surface/adsorbed molecule interface.

Complete modeling of a molecule chemisorbed on a rough metal
surface requires multi-scale simulations, in which the electronic
structure of a molecule and its local environment are treated by
quantum chemical methods while a larger-scale environment is
accounted for by a mean-field approximation. However, to
understand the chemical bonding effects in SERS, it is enough to
simulate only the local environment of a molecule. This is
consistent with the ``adatom model'', \cite{Moskovits_Rev} which
assumes that the atomic-scale roughness features determine the hot
spots on a metal surface. Obviously, such an approach does not
account for the electromagnetic enhancement due to the excitation
of surface plasmons in the metal. An interference between the
chemical and the electromagnetic effects is ignored as well. To
get the total enhancement factor one has to use a conventional
phenomenological relation \cite{ChemEnh2} which assumes that both
effects enter in multiplicative fashion. In most previous
theoretical studies of SERS, only a few metal atoms were used to
model the molecule/surface interaction. For instance, in
Ref.~\cite{Aroca} the authors considered a complex with a single
Ag atom to mimic a chemisorption of phthalimide on a silver
surface and obtained a reasonable agreement with experiments.
Recently, Schatz and co-workers reported a resonance Raman
response calculation procedure using phenomenological lifetime
parameters for electronic polarizability derivatives.
\cite{SchatzRR} This procedure has been applied to study SERS of
pyridine adsorbed on silver clusters up to a few tens of atoms.
\cite{Schatz_JACS, Jensen07,Jensen09} The authors identified three
different mechanisms contributing to SERS: change in static
polarizability of the molecule upon adsorption, resonant
enhancement due to charge--transfer transitions, and
electromagnetic enhancement due to coupling with a strong
excitation in a metal cluster, and discussed their role in SERS.
In Ref.~\cite{Jensen09} some examples of different molecules and
pyridine with substituents have been considered. However, no
systematic studies of other molecular structures have been done
yet. Given the differences in the chemical bonding of pyridine and
thiols to metal surfaces, it is important to know how general
these findings are and whether they can be applied to other types
of metal--molecular coupling or to other molecules.

In this paper, we address effects related to chemical bonding on
Raman scattering from benzenethiol adsorbed on silver clusters
Ag$_n$, $n = 6-11$, which is one of the most commonly used
analytes in SERS experiments. We link the enhancement of Raman
response to the modification of the molecular electronic structure
due to adsorption. Excitation energies and transition moments of
the metal--molecular complexes are substantially different from
those of bare metal clusters and of isolated benzenethiol. For
off-resonant excitations, Raman response is enhanced by one order
of magnitude. To understand the importance of the cluster
geometry, we modeled a pool of 7 different metal--molecular
complexes. We identify the effects that contribute to the
off-resonant Raman enhancement, such as the \emph{molecular
orientational effect}, the effect of the \emph{local symmetry} of
the adsorbate, and the effect of the \emph{proximity} of the
vibrational mode to a binding site. Stochastic modulation of these
effects by thermal motion is related to ``blinking'' events
observed in recent SERS experiments on 4-aminobenzenethiol.
\cite{Moerner,Natelson} We compare the computed vibrational
frequency shifts and relative Raman scattering cross sections of
the the metal--molecular complexes to experimental SERS data for
benzenethiol adsorbed on silver surfaces.
\cite{VanDuyne09,Carron,SERS-Roshi,VanDuyne05}

Our theoretical approach to compute Raman scattering cross section
is based on analytical derivatives of frequency-dependent
polarizabilities. \cite{Rappoport} The frequency dependence of
polarizability derivatives includes resonance enhancement effects
in an approximate fashion, which allows us to describe both
off-resonance and resonance enhancement effects on an equal
footing and to study resonance excitation profiles of Raman
scattering cross sections. However, this approach does not take
into account the finite lifetime of the excited states. As a
result, the Raman scattering cross sections computed within our
approach diverge in the strictly resonant case.

We demonstrate that formation of mixed metal--molecular electronic
states results in a resonance structure in the Raman excitation
profiles (REPs) for molecular vibrations within the excitation
range $1.6-3.0$~eV, which is far below the purely intramolecular
transitions. Our lower-bound estimates of the Raman signal
enhancement due to chemical bonding is of the order of $10^2-10^3$
in the absence of intramolecular resonances. The chemical bonding
effect thus provides additional flexibility in controlling the
enhancement of Raman scattering from chemisorbed species.

The paper is organized as follows: in Sec.~\ref{sec:comp-details}
we describe the details of our computational procedure. In
Sec.~\ref{sec:geometries}-\ref{subsec:modes} the computed results
are presented and different mechanisms contributing to the Raman
enhancement are discussed. In particular, we show that the
orientation of the benzene ring has a substantial effect on the
Raman scattering cross section. We also identify metal-molecular
states responsible for the resonant enhancement of the Raman
response of the computed complexes. In Sec.~\ref{subsec:discuss}
the computed Raman spectra are compared to available experimental
data. The conclusions are formulated in
Sec.~\ref{sec:conclusions}.

\section{Computational Details}
\label{sec:comp-details}

Density functional calculations were performed using the quantum
chemistry package Turbomole, version 5.10. \cite{TM_general}
Triple-$\zeta$ valence-polarization basis sets (def2-TZVP
\cite{BS_def2}) were used for the main group elements while for
silver atoms we employed split-valence basis sets with
polarization (def2-SV(P) \cite{BS_defSVP}) and effective core
potentials (ECP) comprising the 28 core electrons and accounting
for scalar relativistic effects. \cite{BS_ecp} Preliminary
calculations showed that expanding the molecular basis set up to
quadruple-$\zeta$ quality \cite{BS_def2} does not significantly
alter the results.

Our choice of the computed exchange-correlation functional was
guided by the balance between the accuracy of the computed Raman
scattering cross sections and an accurate description of
electronic and geometric parameters of Ag$_n$ clusters. We chose
the hybrid functional of Perdew, Burke, and Ernzerhof (PBE0)
\cite{PBE0} which provides good accuracy for frequency-dependent
polarizabilities and Raman scattering cross sections.
\cite{Caillie_pol,Caillie_raman} While many computational studies
of metal clusters have been traditionally performed with
gradient-corrected exchange-correlation functionals,
\cite{Ag_abs,Ag_abs_PRB} hybrid functionals were found to yield
good structural parameters and band gaps in extended systems.
\cite{Kresse}

The charge-transfer error of approximate exchange-correlation
functionals is a significant challenge in TDDFT calculations,
although several recent works have proposed approaches to address
this problem. \cite{Neepa,CAMB3LYP,Troy,FuncCT} The
charge-transfer error results in an overestimation of electronic
polarizabilities and an underestimation of electronic excitation
energies, however these effects are ameliorated by inclusion of a
fraction of non-local Hartree--Fock exchange in hybrid functionals
such as PBE0, \cite{Caillie_raman,CAMB3LYP} which we use in this work.

Ground state structure optimizations of silver clusters Ag$_n$
with $n$ = 6--11 were performed using the PBE0 functional.
Force-constant calculations were used to confirm that the
optimized structures correspond to local minima of the electronic
potential energy surface. \cite{TM_force} Many energetically
close-lying structure minima are known to exist in small Ag$_n$
clusters \cite{Ag_abs, Idrobo, Idrobo11}. We concentrated on
structures suitable to represent atomic-scale roughness features,
especially those with pyramidal or bipyramidal shape. For
instance, for the Ag$_6$ cluster the pyramidal $C_{5v}$ isomer,
which might be considered as a more realistic model for
atomic-scale roughness, rather than the lowest-energy planar
structure of $D_{3h}$ symmetry \cite{Ag_abs, Idrobo} was used. For
each optimized cluster structure, the 200 lowest electronic
excitations were computed using TDDFT in the frequency-based
linear response regime.\cite{Casida,TM_excitedBauer, TM_excited}
The obtained line spectra were broadened using an empirical
Gaussian broadening parameter of 0.05~eV.

Oxidative attachment of the benzenethiol molecule
\cite{Whitesides, Maddix} was modeled by placing the benzenethiol
radical (PhS) at different atomic binding sites and  fully
reoptimizing the structures of the PhS--Ag$_n$ complexes. Raman
spectra of benzenethiol and PhS--Ag$_n$ complexes were obtained
from analytical derivatives of frequency-dependent
polarizabilities. \cite{Rappoport} This approach is based on the
polarizability Lagrangian and allows for efficient computation of
Raman spectra of medium-sized and large molecules. Raman spectra
of benzenethiol and PhS--Ag$_n$ complexes were computed for a
low-energy excitation of 0.62 eV (2000 nm) to investigate the
off-resonance Raman scattering enhancement. Raman excitation
profiles of benzenethiol and PhS--Ag$_n$ were computed for the
range of 1.6--3.0 eV for the four strongest vibrational modes of
the benzenethiolate group. No scaling of vibrational frequencies
was applied. The Raman spectra were simulated by Lorentzian
broadening of the line spectra using an empirical linewidth of 5
cm$^{-1}$. A scattering angle of 90 degrees and perpendicular
polarization of both incident and scattered radiation was assumed
unless specified otherwise.

\section{Results and Discussion}
\label{sec:results}

\subsection{Silver Clusters and Metal-Molecular Complexes}
\label{sec:geometries}

The computed structures of the clusters Ag$_n$, $n=6-11$, are in
agreement with results of previous works. \cite{Ag_abs} The
clusters with an even number of atoms $n$ were found to be
closed-shell singlets while the odd-numbered clusters are spin
doublets. The relative energies of the structures are dependent on
the particular choice of the exchange-correlation functional.
Among the simulated structures, several metal clusters were
selected based on their similarity in geometry. In particular, the
geometries derived from the bipyramidal structure of Ag$_7$  were
chosen.

\begin{figure}
  \caption{Optimized geometries of
PhS--Ag$_n$, $n=6-11$ complexes that were utilized in the study.
For $n=7$ two isomers with different types of binding for the
molecule were considered.}
  \label{fig:geometry}
  \begin{center}
   \includegraphics[width=0.5\linewidth, clip=true]{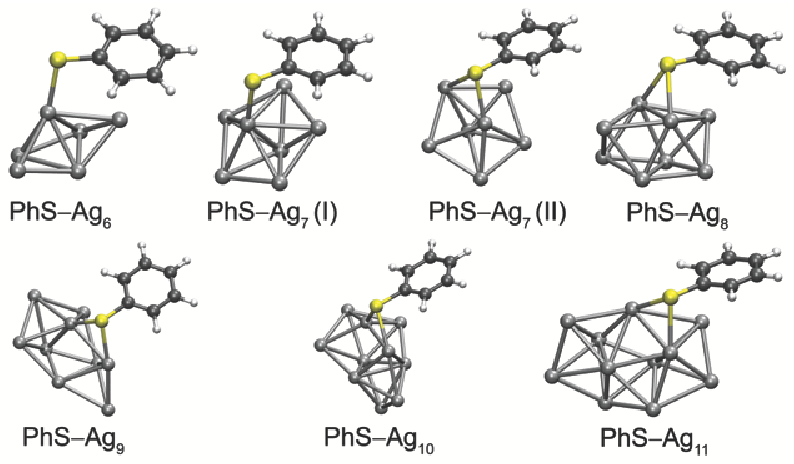}
  \end{center}
\end{figure}

The benzenethiol molecule binds to the Ag metal surface via the
thiolate bond. To construct the metal--molecular complexes the
central atom of the bipyramid has been chosen as a binding site.
The final optimized geometries of PhS--Ag$_n$, $n=6-11$, are shown
in Fig.~\ref{fig:geometry}. In most configurations the sulfur atom
migrates to the middle of an Ag--Ag bond forming a bridging motif.
This is consistent with previous theoretical studies of
benzenethiol chemisorbed on Au surfaces. \cite{PhS_Au_geom} For a
better understanding of properties of the PhS--Ag$_n$ complexes
the simplest structure---benzenethiol bound to a single Ag
atom---has been simulated. Addition of benzenethiol to the
$C_{5v}$--symmetric Ag$_6$ cluster leads  to a rearrangement of
the metal atoms to form a $C_{2v}$--symmetric cluster structure,
see Fig.~\ref{fig:geometry}. However, we find that the $C_{2v}$
structure of Ag$_6$ does not correspond to a local minimum for the
PBE0 functional. Starting with approximately similar initial
conditions, two stable isomers of the PhS--Ag$_7$ complex with an
on-top and a bridging binding motif have been obtained,
respectively. The energy difference between these two structures
is approximately 3~kcal/mol, with the lower configuration
corresponding to the complex with the bridging motif.

The structural characteristics of the PhS--Ag$_n$, $n=1,6-11$,
complexes are given in Tab.~\ref{tbl:geometry}. They include the
lengths of the Ag--S and the S--C bonds, the angle~C--S--Ag (for
the bridging motif the calculated angle is between the S--C bond
and a shortest line connecting the sulfur atom with the nearest
Ag--Ag bond), and the distance between the carbon atom adjacent to
the S--C bond and the closest silver atom. The latter parameter
characterizes the shortest non-bonding distance between the
benzene ring and the Ag$_n$ cluster. The binding energies between
benzenethiol and silver clusters were estimated as
\begin{equation}
E_b(\mbox{PhS--Ag}_n) =
E(\mbox{Ag}_n)+E(\mbox{PhSH})-E(\mbox{PhS--Ag}_n)-\frac{1}{2}
E(\mbox{H}_2),
\label{binding_E}
\end{equation}
neglecting the basis set superposition errors. \cite{note_err} The
variation of the Ag--S and C--S bond lengths for the complexes
with the same type of binding is less than 4~\%. The angle between
the C--S and Ag--S bonds is consistent with the experimental data
obtained for benzenethiol adsorbed on Au (111). \cite{Whelan}
There are no obvious correlations of the structure parameters with
the size of the Ag$_n$ clusters. The binding energies vary
substantially for the complexes. PhS--Ag$_n$ complexes with odd
$n$ form closed-shell electronic configurations and have larger
binding energies.
\begin{table}
\begin{center}
\caption{The computed structural properties of PhS--Ag$_n$,
$n=1,6-11$, complexes. The lengths of the Ag--S and the C--S
bonds, and the shortest distance between the aromatic ring and the
cluster, C$_2$--Ag, are given in \aa. The C$_2$ denotes a carbon
atom adjacent to the C--S bond. The angle C--S--Ag is given in
degrees and the binding energy of a molecule and a cluster is
calculated according to Eq.~\ref{binding_E} in
kcal/mol.}\label{tbl:geometry}
\begin{tabular}{|l|c|c|c|c|c|c|}
\hline
Complex & Binding & Ag--S &  C--S & $\angle$ C--S--Ag &  C$_2$--Ag & $E_b$ \\
\hline
PhS--Ag$_1$ & on-top & 2.339        & 1.769 & 102.830 & 3.975 & 12.34 \\
PhS--Ag$_6$ & on-top & 2.389        & 1.756 & 101.470  & 2.454 & 7.56 \\
PhS--Ag$_7$ (I) & on-top & 2.421        & 1.757 & 96.606  & 2.669 &17.71\\
PhS--Ag$_7$ (II) & bridge & 2.521/2.589 & 1.768 & 103.673 & 2.629 &20.67\\
PhS--Ag$_8$ & bridge & 2.573/2.587 & 1.767 & 109.449 & 2.600 &3.31\\
PhS--Ag$_9$ & bridge & 2.429/2.458 & 1.774 & 109.682 & 3.646 &25.06\\
PhS--Ag$_{10}$ & bridge & 2.480/2.495 & 1.769 & 108.946 & 3.684 &15.37\\
PhS--Ag$_{11}$ & bridge & 2.544/2.546 & 1.768 & 105.044 & 2.523 &23.32\\
\hline
\end{tabular}
\end{center}
\end{table}

Two types of benzene ring orientation are distinguishable. In the
complexes PhS--Ag$_n$, $n=6,7,8,11$ the benzene ring oriented
towards the metal cluster, allowing several non-bonding C--Ag
interactions. In the complexes PhS--Ag$_n$, $n=9,10$ the benzene
ring is directed outward and the distance between the ring and the
Ag$_n$ cluster is about 40~\% longer than in the previous type.

In the following sections, we provide a detailed analysis of the
computed optical and Raman spectra of PhS--Ag$_n$ complexes using
representative examples of the aforementioned binding structures
and molecule orientations. The spectra of the other complexes are
provided in the Supplementary Information.

\subsection{Electronic Excitation Spectra.}
\label{subsec:el-exc}

\begin{figure}
  \begin{center}
   \includegraphics[width=0.5\linewidth, clip=true]{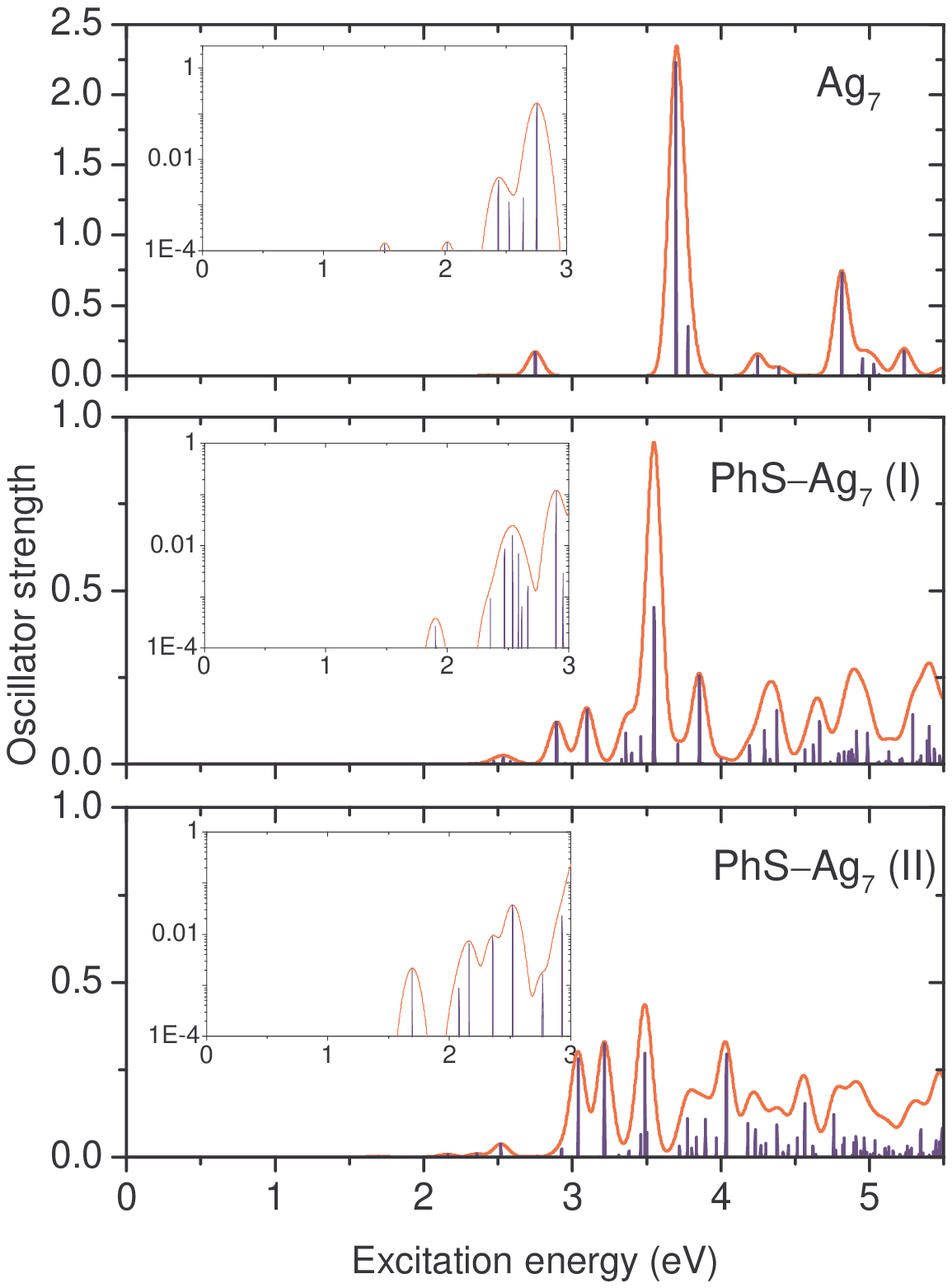}
\caption{Electronic excitation spectra of Ag$_7$ cluster and two
isomers of the PhS--Ag$_7$ complex with different types of binding
on-top and bridge. The insets show the low-frequency parts of the
spectra in logarithmic scale for the range $1.0-3.0$~eV. }
  \label{fig:excitation}
  \end{center}
\end{figure}

The electronic excitations of neat benzenethiol are in the
ultraviolet spectral range. The computed energies of the two
lowest $a'$ transitions, 4.87~eV and 5.37~eV, are in good
agreement with the measured values of 4.42~eV and 5.25~eV for
benzenethiol in a solution, respectively. \cite{BTexp} The
simulated absorption spectra of bare silver clusters are
consistent with the experimental data from Ref.~\cite{Ag_abs}. In
the energy range of 3--5~eV, the clusters of higher symmetry, such
as Ag$_7$ ($D_{5h}$) and Ag$_{10}$ ($D_{2d}$), have few strong
transitions with oscillator strengths higher than~1. The
low-frequency parts of the absorption spectra overlap with the
visible range and some extend down to 0.5~eV. They consist of weak
transitions with oscillator strengths in the range
$10^{-1}-10^{-4}$. As an example, the electronic excitation
spectra of the Ag$_7$ cluster and two isomeric complexes
PhS--Ag$_7$ (I) and (II) are shown in Fig.~\ref{fig:excitation}.
The spectra of the complexes differ substantially from the spectra
of the corresponding silver clusters. In contrast to
silver--pyridine complexes, \cite{Jensen07} the strong electronic
transitions are quenched in most of the studied PhS--Ag$_n$
structures. In the cases where these excitations are still
identifiable, we observe a red-shift of the order of $0.1-0.2$~eV,
see Fig.~\ref{fig:excitation}, except for the PhS--Ag$_{10}$
complex where the transition is blue-shifted by about 0.1~eV.

\subsection{Off-Resonance Raman Spectra}
\label{subsec:offres}

\begin{figure}
  \begin{center}
   \includegraphics[width=0.5\linewidth, clip=true]{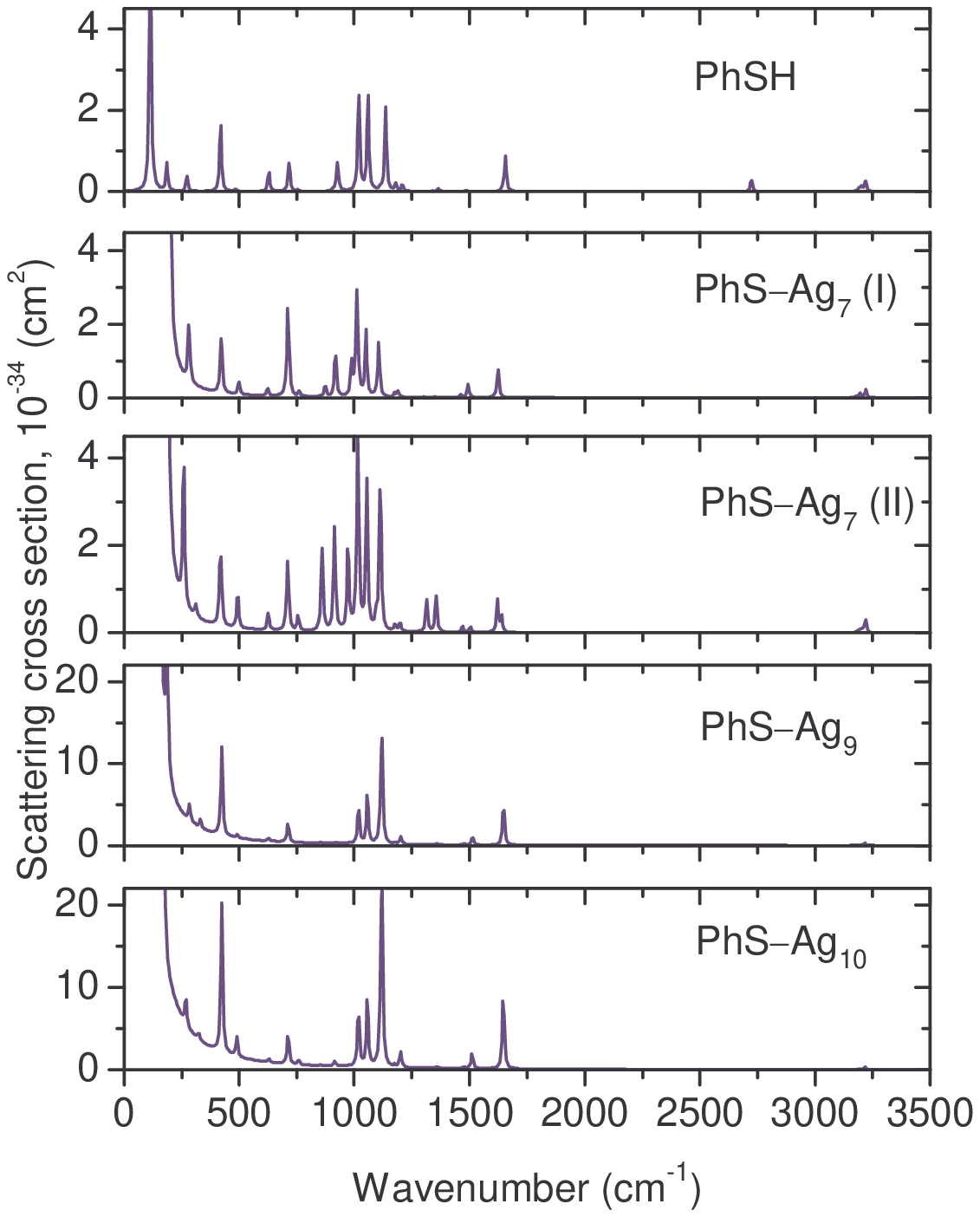}
  \caption{The computed Raman spectra of
isolated benzenethiol (PhSH) and four different complexes
PhS--Ag$_7$ (I) and (II), PhS--Ag$_9$, and PhS--Ag$_{10}$ for the
excitation energy 0.62~eV (2000~nm), which is below all electronic
transitions in the complexes.}
    \label{fig:bt-raman-2000}
  \end{center}
\end{figure}

Off-resonance Raman spectra of benzenethiol and of PhS--Ag$_n$,
$n=6-11$, complexes were simulated using an excitation energy of
0.62~eV (2000~nm), which is below the lowest electronic transition
in these systems. Three different quantities have been computed:
Raman cross sections for characteristic molecular vibrational
modes; the total scattering cross section of the complex, and the
integrated cross section with a low-frequency cutoff at 200
cm$^{-1}$. The latter quantity excludes almost all vibrational
modes of the metal cluster. Most important for SERS applications
are molecular vibrational modes about 1000~cm$^{-1}$. Enhancement
of the scattering cross sections of these modes is used to
characterize the surface enhancement. However, the integrated
cross sections are introduced to analyze the modulations of the
Raman response rather than for estimation of the enhancement
factor. The computed Raman spectra are given in
Fig.~\ref{fig:bt-raman-2000}.

The computed Raman spectrum of isolated benzenethiol is in
reasonable  agreement with the experimental off-resonant spectrum
of neat benzenethiol recorded for a 785~nm excitation.
\cite{VanDuyne09} The dominant vibrational modes in the Raman
spectrum of benzenethiol are totally symmetric and belong to the
$a'$ irreducible representation of the $C_s$ point group. The
computed vibrational band at $\omega_1=1019$~cm$^{-1}$ is assigned
to the ring breathing mode, and the band at
$\omega_2=1059$~cm$^{-1}$ corresponds to a ring deformation mode.
The C--S stretching mode is computed at $\omega_3=1136$~cm$^{-1}$
while the computed Raman band at $\omega_4=1656$~cm$^{-1}$ is
associated with a totally symmetric ring stretching mode. Our
results are consistent with previous assignments.
\cite{PhSH-Green} The computed vibrational frequencies of
benzenethiol are blue-shifted by 10--50~cm$^{-1}$ with respect to
experimental data due to incomplete basis sets and neglect of
anharmonicity. The totally symmetric modes assigned above
correspond to the experimentally measured vibrational frequencies
$\omega_1^{\rm exp}=1004$~cm$^{-1}$, $\omega_2^{\rm
exp}=1027$~cm$^{-1}$, $\omega_3^{\rm exp}=1094$~cm$^{-1}$, and
$\omega_4^{\rm exp}=1583$~cm$^{-1}$, respectively, for neat
benzenethiol. \cite{VanDuyne09} Comparison of computed Raman
intensities of isolated benzenethiol to experimental data is
provided in the Supplementary Information.
\begin{table}
\begin{center}
\caption{Isotropic polarizabilities $\alpha_{\mbox{iso}}$ (a.u.),
HOMO-LUMO gaps (eV), Raman cross sections for the
$\omega_4=1583$~cm$^{-1}$ vibrational mode $\sigma_{4}$ (cm$^2$),
integrated Raman scattering cross sections $\sigma_{\mbox{tot}}$
(cm$^2$), and fractions $w$ (\%) of Raman scattering cross section
from modes with frequencies larger than 200 cm$^{-1}$ for
benzenethiol (PhSH) and PhS--Ag$_n$ complexes. The excitation
wavelength is 2000 nm.} \label{table:cross-sect}
\begin{tabular}{|l|c|c|c|c|c|}
\hline Complex & $\alpha_{\mbox{iso}}$ & $E_{\rm gap}$ &
$\sigma_{4}\dot 10^{-33}$ & $\sigma_{\mbox{tot}}$ &
$w$ \\
\hline
PhSH            &  85.8  & 5.96 &  1.39 & 3.34$\cdot10^{-32}$ & 64.0 \\
PhS--Ag$_1$     & 130.7  & 3.35 & 16.50 & 1.53$\cdot10^{-30}$ & 10.2 \\
PhS--Ag$_6$     & 351.7  & 1.53 &  1.63 & 2.16$\cdot10^{-30}$ &  2.8 \\
PhS--Ag$_7$ (I)  & 365.2  & 2.80 &  1.35 & 1.04$\cdot10^{-30}$ &  2.7 \\
PhS--Ag$_7$ (II) & 374.9  & 2.54 &  0.64 & 1.23$\cdot10^{-30}$ &  4.8 \\
PhS--Ag$_8$     & 425.6  & 1.34 &  0.48 & 1.67$\cdot10^{-30}$ &  4.3 \\
PhS--Ag$_9$     & 478.2  & 2.09 &  7.98 & 1.35$\cdot10^{-29}$ &  0.6 \\
PhS--Ag$_{10}$  & 529.3  & 1.22 & 14.43 & 2.28$\cdot10^{-29}$ &  0.6 \\
PhS--Ag$_{11}$  & 568.8  & 1.95 &  1.00 & 1.81$\cdot10^{-30}$ &  4.0 \\
\hline
\end{tabular}
\end{center}
\end{table}

 Binding of benzenethiol to Ag$_n$ clusters leads to an
overall increase of off-resonant Raman scattering cross sections
and to a significant redistribution of band intensities in the
Raman spectrum. This effect is known as the \emph{static
enhancement } mechanism of SERS. \cite{Moskovits_Rev} This effect
has been attributed to the decrease of the energy gap between the
highest occupied molecular orbital (HOMO) and the lowest
unoccupied molecular orbital (LUMO) and the ensuing enhancement of
the electronic polarizability and its derivatives. \cite{Jensen09}
The Raman cross section of the molecular vibrational modes
$\omega_{1-4}$ in the computed metal-molecular complexes can be up
to 10 times larger than that in isolated benzenethiol. This value
is comparable to the results of Ref.~\cite{Schatz_JACS} for
pyridine. However, we observe a strong dependence of the
enhancement factor on the geometry of a metal cluster and the
orientation of the molecule. For example, the Raman signal from
the modes $\omega_{1-4}$ in PhS--Ag$_7$ (I and II) complexes is
almost unenhanced compared to PhSH, see
Fig.~\ref{fig:bt-raman-2000}. Also, Raman signals from different
molecular modes have different enhancement factors. To
characterize the off-resonant enhancement of Raman scattering, the
cross sections of the $\omega_4$ vibrational mode have been used.
Tab.~\ref{table:cross-sect} shows the computed off-resonant
electronic polarizabilities, HOMO--LUMO gaps, Raman cross sections
corresponding to the $\omega_4$ vibrational mode, and integrated
Raman scattering cross sections of benzenethiol and PhS--Ag$_n$
complexes for the excitation energy of 0.62~eV (2000~nm). The
integrated cross sections are computed by summation of Raman
scattering cross sections over all vibrational modes of the
complex, including the vibrations of the metal cluster. The
electronic polarizabilities are to a good approximation additive
and thus increase linearly with the number of Ag atoms. The
integrated Raman scattering cross sections of PhS--Ag$_n$
complexes are larger than that of benzenethiol by 2--3 orders of
magnitude but do not increase monotonically with the cluster size
$n$. Molecular vibrational modes contribute only a small fraction
(less than 5~\% with the exception of PhS--Ag, see
Tab.~\ref{table:cross-sect}) to the integrated Raman scattering
cross section while the largest part of the scattered radiation
arises from vibrations within the metal cluster.

{\bf Orientation effect.} We find a significant dependence on the
orientation of the benzene ring: the largest integrated Raman
scattering cross sections are computed for the complexes
PhS--Ag$_n$, $n=9,10$, in which the benzene ring points outward
from the cluster surface, see Fig.~\ref{fig:geometry}. In
contrast, in the PhS--Ag$_n$, $n=6-8,11$ complexes the aromatic
ring is oriented towards the metal cluster and non-binding C--Ag
interactions reduce the integrated Raman scattering cross sections
by about one order of magnitude. This orientation effect appears
to be the major influence in the off-resonance Raman spectra of
PhS--Ag$_n$ complexes, while the effect of increasing electronic
polarizability and the characteristic alternation of HOMO--LUMO
gaps in between even- and odd-numbered metal clusters Ag$_n$ seem
to be less important. The non-binding interactions of the benzene
ring with the Ag$_n$ cluster in the PhS--Ag$_n$, $n=6-8,11$
complexes result in an efficient quenching of Raman scattering
from metal cluster modes, increasing the relative weight of
molecular vibrations.

{\bf Symmetry effect.} The relative intensities of vibrational
modes in Raman spectra of
PhS--Ag$_n$ complexes might be explained based on their symmetry
and proximity to the Ag$_n$ cluster. Comparison of the Raman
spectra of the isomers PhS--Ag$_7$ (I) and PhS--Ag$_7$ (II)
exemplifies the effect of the local symmetry of the benzene ring
on the off-resonant enhancement. The PhS--Ag$_7$ (I) isomer is,
similar to the isolated benzenethiol molecule, $C_s$-symmetric
while the local symmetry of the benzene ring in the PhS-Ag$_7$
(II) isomer is perturbed by unsymmetric bonding to the cluster.
The symmetry of the benzene ring affects the aromatic C--C stretching bands
in the region of 1620--1660~cm$^{-1}$. Only the totally symmetric
aromatic C--C stretching vibration ($\omega_4=1656$~cm$^{-1}$ in
benzenethiol, denoted as $8a$ in Wilson's notation
\cite{Benzene-Wilson}) is observed in the Raman spectra of PhSH
and PhS--Ag$_7$ (I) while the corresponding non-totally symmetric
vibration at lower frequency (1641~cm$^{-1}$ in benzenethiol, $8b$
in Wilson's notation) is at least one order of magnitude weaker
and not observed in the experiment. The lower symmetry of the
PhS--Ag$_7$ (II) manifests itself in the redistribution of Raman
scattering cross sections among the aromatic C--C stretching modes such that
the lower-frequency $8b$ mode gains intensity. As a consequence, a
doublet of vibrational bands is formed in the aromatic C--C stretching
region. In the PhS--Ag$_n$, $n=9,10$, complexes, the interaction
between the benzene ring and the Ag$_n$ cluster is weak and local
$C_s$ symmetry of the benzene ring is essentially unperturbed.
This results in only one observable aromatic C--C stretching band in the
Raman spectra of these complexes.

{\bf Proximity effect.} The proximity effect on the Raman
scattering cross section may be observed in the region between
1000--1150~cm$^{-1}$, including ring deformation modes at
$\omega_1=1019$~cm$^{-1}$ and $\omega_2=1059$~cm$^{-1}$ and the
C--S stretching mode at $\omega_3=1136$~cm$^{-1}$ in benzenethiol,
respectively. In both PhS--Ag$_7$ (I) and PhS--Ag$_7$ (II)
clusters, the ring breathing mode at 1011--1015~cm$^{-1}$ and the
ring deformation mode at 1050--1055~cm$^{-1}$ are strongly
enhanced by the interaction with the Ag$_n$ clusters. With benzene
rings further away from the Ag$_n$ clusters in PhS--Ag$_n$,
$n=9,10$, enhancement of ring deformation modes is diminished in
these clusters, while the C--S stretching mode is more pronounced.

All three effects governing off-resonant surface enhancement of
Raman response of PhS--Ag$_n$ complexes, namely the relative
orientation of the benzene ring with respect to the cluster, the
local symmetry of the benzene ring, and the proximity of the
particular vibrational mode to the binding site, are interrelated.
Their modulation by thermal large-amplitude motion of the aromatic
ring may serve as a simple and plausible explanation for the
``blinking'' events observed in SERS experiments on
4-aminobenzenethiol attached to Au bowtie nanoantennas
\cite{Moerner} or to molecular junctions. \cite{Natelson}

\subsection{Raman Excitation Profiles}
\label{subsec:modes}

Raman excitation profiles of four strong totally symmetric
benzenethiol vibrational modes,  $\omega_1 = 1019$~cm$^{-1}$,
$\omega_2 = 1059$~cm$^{-1}$, $\omega_3 = 1136$~cm$^{-1}$, and
$\omega_4 = 1656$~cm$^{-1}$, were calculated for the excitation
energy range $1.6 - 3.0$~eV ($413-775$~nm). These modes are
dominant in experimental SERS spectra of benzenethiol.
\cite{VanDuyne09, Carron}

\begin{figure}
  \begin{center}
   \includegraphics[width=0.5\linewidth, clip=true]{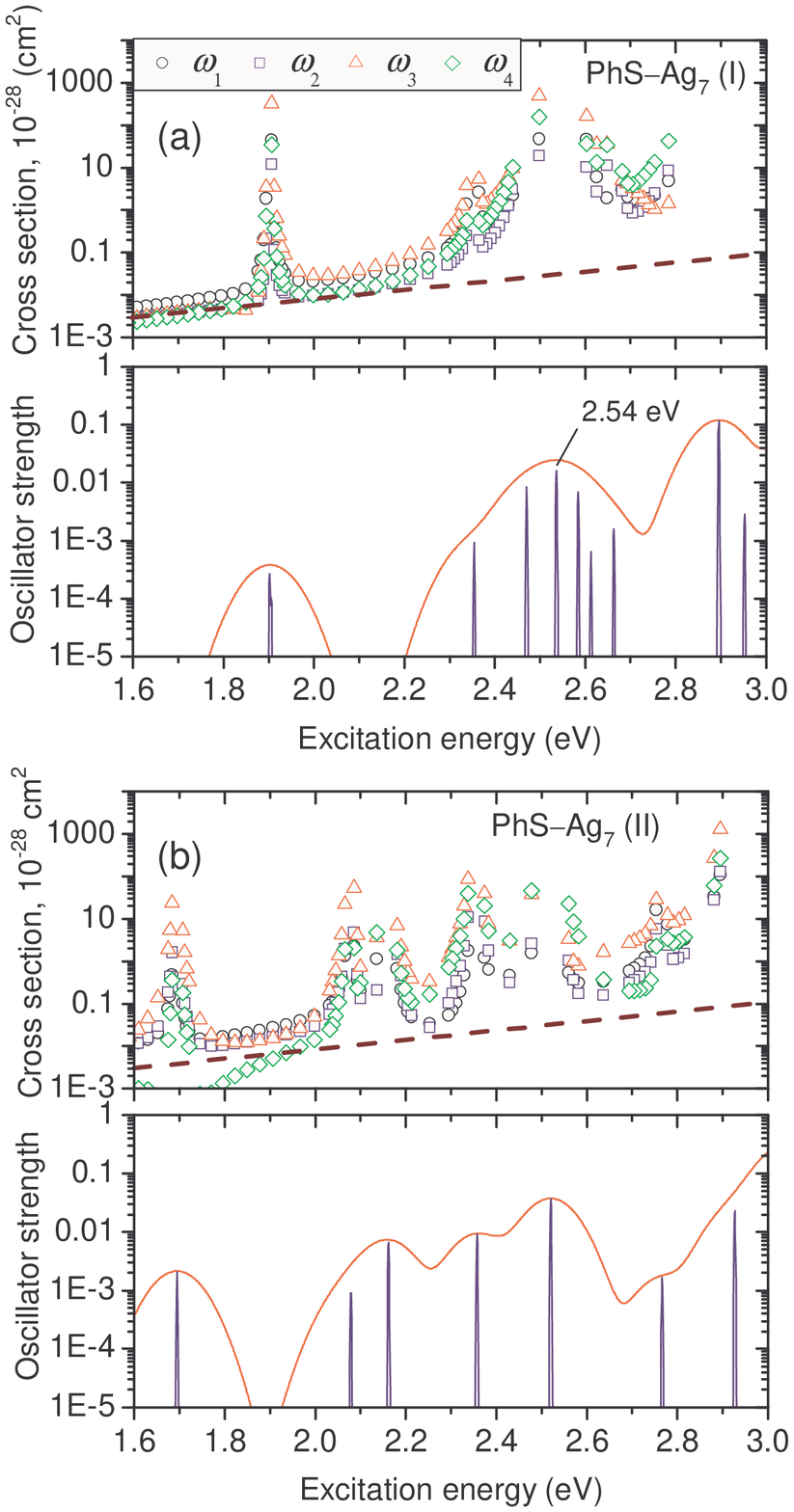}
  \caption{The Raman excitation profiles
(REPs) together with the electronic excitation spectra for (a) the
PhS--Ag$_7$~(I) and (b) the PhS--Ag$_7$~(II) complexes. The
vibrational modes are $\omega_1=1019$~cm$^{-1}$,
$\omega_2=1059$~cm$^{-1}$, $\omega_3=1136$~cm$^{-1}$, and
$\omega_4=1656$~cm$^{-1}$. The dashed tilted line in the REPs
plots corresponds to the response of isolated benzenethiol
averaged over all four modes.}\label{fig1:rep}
  \end{center}
\end{figure}

\begin{figure}
  \begin{center}
   \includegraphics[width=0.5\linewidth, clip=true]{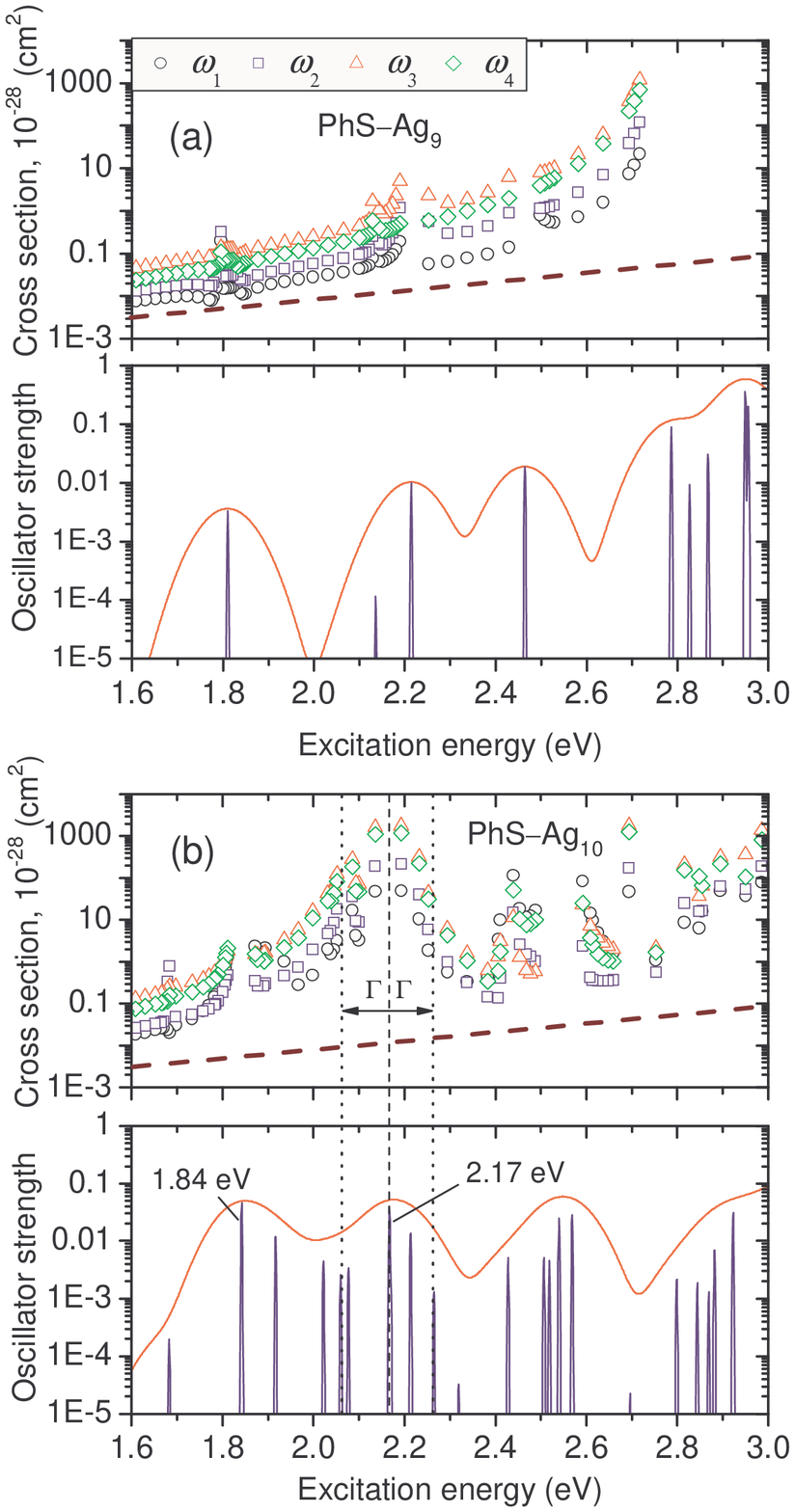}
  \caption{The Raman excitation profiles
together with the electronic excitation spectra for (a) the
PhS--Ag$_9$ complex and (b) the PhS--Ag$_{10}$ complex. The
vibrational modes are $\omega_1=1019$~cm$^{-1}$,
$\omega_2=1059$~cm$^{-1}$, $\omega_3=1136$~cm$^{-1}$, and
$\omega_4=1656$~cm$^{-1}$. The dashed tilted line in the REPs
plots corresponds to the response of isolated benzenethiol
averaged over all four modes. $\Gamma = 0.1$~eV is a
phenomenological parameter characterizing homogeneous linewidth of
electronic states. }\label{fig2:rep}
  \end{center}
\end{figure}

In Figs.~\ref{fig1:rep} and ~\ref{fig2:rep} we provide REPs and
electronic excitation spectra of four different complexes
PhS--Ag$_7$ (I), PhS--Ag$_7$ (II), PhS--Ag$_9$, and PhS--Ag$_{10}$
in logarithmic scale. For comparison we also show REP of isolated
benzenethiol averaged over all four modes. The simulated
scattering cross sections for the vibrational modes $\omega_{1-4}$
in isolated benzenethiol differ by no more than a factor of two,
therefore an averaged value of REP of benzenethiol (dashed line)
is used to simplify the plots. The slope of the REP of
benzenethiol is close to 4 and stems from the $\omega^4$
dependence of Raman scattering cross sections. \cite{Long} The
effect of intramolecular resonances is small in the studied range
of excitation energies.

The isomers PhS--Ag$_7$ (I) and (II) in Fig.~\ref{fig1:rep}
represent two distinct bonding patterns (on-top and bridge),
whereas the structure of the silver cluster remains essentially
the same. The electronic excitation spectra of the complexes are
quite dissimilar, which reflects the sensitivity of the electronic
states to the binding pattern. For example, the lowest electronic
excitation of the PhS--Ag$_7$ (II) complex is 0.2~eV lower in
energy than that of the PhS--Ag$_7$ (I) complex. All transitions
within the excitation range $1.6 - 3.0$~eV are weak with
oscillator strengths $10^{-1} - 10^{-4}$.

{\bf Cluster--molecule resonant effect}. The REPs illustrate the
frequency dependence of SERS and allow for identification of two
different contributions to chemical Raman enhancement. These are
the baseline shifts of the REPs of the complexes compared to that
of isolated benzenethiol and the additional modulation by
resonance-type features. The baseline shift is attributed to the
off-resonant enhancement discussed in the previous section while
the resonances arise from electronic transitions between
metal--molecular states. This is consistent with the conventional
picture of chemical enhancement.  \cite{Schatz_Rev} The
off-resonant enhancement factor for both isomers of the
PhS--Ag$_7$ complex is of the order of 1 and can be explained by
the orientational effect discussed in the previous section.

The relative enhancement for different modes depends on the
excitation energy. For example, the $\omega_1$ vibrational mode
has the largest enhancement for the PhS--Ag$_7$ (I) complex in the
energy range below the first electronic transition, whereas the
$\omega_3$ becomes the strongest Raman mode above it. This
observation is in accord with the empirical Tsuboi rule
\cite{Tsuboi}: the lowest electronic excitation in the complex
PhS--Ag$_7$ (I) involves the Ag--S bond and has a larger impact on
the C--S stretching mode ($\omega_3$).

The structures of the PhS--Ag$_9$ and PhS--Ag$_{10}$ complexes
differ only by a single Ag atom, which is located approximately
$4.8$~$\AA$ from the binding sites. The PhS--Ag$_9$ complex has a
closed-shell electronic structure and shows a smaller number of
electronic transitions as compared to the PhS--Ag$_{10}$. Both
structures show 10--100 fold enhancements of Raman signals outside
of resonant regions. Enhancement factors of different vibrational
modes differ by about one order of magnitude. This can be seen by
comparing the Raman cross sections of the complexes to that of
isolated benzenethiol for the excitation energy 1.6~eV, Fig.
\ref{fig2:rep}.

\begin{figure}
  \begin{center}
   \includegraphics[width=0.3\linewidth, clip=true]{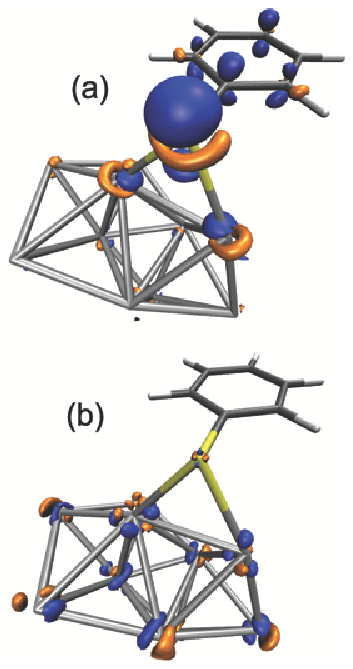}
  \caption{The transition electron densities
for (a) the 2.17~eV electronic excitation and (b) the 1.84~eV
electronic excitation in the PhS--Ag$_{10}$ complex. The former
transition results in the strong resonant features in the REP,
whereas the latter one has much weaker effect. The light grey
(orange) color corresponds to the transition density of $-0.002$
a.u., the dark grey (blue) corresponds to the transition density
of $0.002$ a.u.}\label{fig3:rep}
  \end{center}
\end{figure}

Most of electronic excitations shown in Figs.~\ref{fig1:rep} and
\ref{fig2:rep} result in resonance-type divergences of the REP.
However, Raman enhancement in pre-resonant regions cannot be
characterized by the oscillator strengths of the respective
electronic transitions only. For instance, in the PhS--Ag$_{10}$
complex the electronic transition at 2.17~eV has an oscillator
strength 0.04, which is smaller than that of the 1.84~eV
transition (oscillator strength 0.05). Nevertheless, the former
excitation induces a large peak in the REP, see
Fig.~\ref{fig2:rep} (b), while the latter has a much smaller
effect. These two excitations differ by the degree to which Ag--S
bonds are affected by the electronic excitations, as can be seen
in Fig.~\ref{fig3:rep}. The 2.17~eV electronic excitation is
localized around the bonding sites and does not involve all atoms
in the Ag$_n$ cluster, whereas the 1.84~eV electronic excitation
is delocalized. This observation lends support to the underlying
assumption of this work that the effects of chemical bonding on
surface enhancement can be modeled to a significant extent using
finite-size metal clusters.

The occurrence of strong resonant Raman peaks is a common feature
of all considered PhS--Ag$_n$ complexes. The strong peak in the
REP of the PhS--Ag$_{7}$ (I) complex is centered around the
2.54~eV electronic excitation, and in the REP of PhS--Ag$_{9}$ a
similar feature can be associated with the region $2.8-3.0$~eV.

Our approach does not include finite-lifetime effects and is thus
not applicable in a strictly resonant case. The enhancement of a
Raman signal at resonances scales inversely proportional to the
fourth power of the homogeneous linewidth $\Gamma$ of the involved
electronic states. \cite{Birke} Since $\Gamma$ enters as a
phenomenological parameter, estimates of the corresponding
enhancement factors are inherently imprecise. To obtain
lower-bound estimates of resonant enhancement factors we consider
the Raman scattering cross section at frequencies $\Omega_i \pm
\Gamma$, where $\Omega_i$ stand for electronic excitation
frequencies. For $\Gamma$ we use a pragmatic value $0.1$~eV, which
is consistent with Ref.~\cite{Schatz_JACS}. The estimated
enhancement of Raman signal near the 2.17~eV electronic transition
in the PhS--Ag$_{10}$ complex is of the order of $10^3$ and varies
about 10 times for different vibrational modes. Similar estimates
are also obtained for other structures.

Usually, in SERS literature the enhancement below intramolecular
excitations is associated with charge-transfer excitations that
have very small oscillator strengths. For example, in
Ref.~\cite{Schatz_JACS} electronic transitions at energies
comparable to those considered here were found in a
pyridine-Ag$_{20}$ complex, with enhancements that were also about
$10^3$. The excitations were referred to as charge-transfer
states. In our system, all electronic transitions in the
excitation range $1.6-3.0$~eV have a mixed metal--molecular
character and can be associated with a partial charge transfer
between the cluster and the molecule. However, most of them have
only a small contribution to the resonance Raman enhancement.
Thus, besides narrow singularities, the computed cross sections
are relatively smooth functions of the excitation energy. For
example, the 1.9~eV excitation in PhS--Ag$_7$ (I) complex, see
Fig.~\ref{fig1:rep}, gives almost no enhancement of the Raman
signal if we take into account the finite linewidth of the
transition. The important excitations, such as the excitation
shown in Fig.~\ref{fig3:rep} (a)., are local to the binding site
and are characterized by the change of the electron density around
the binding sites only. It is important to notice that the
difference in the static dipole between a ground and an excited
state of the complexes for such transitions is rather small. For
example, the change in the static dipole moment for the excitation
of the PhS--Ag$_{10}$ complex depicted in Fig.~\ref{fig3:rep} (a)
is ~3.7 Debye.

\subsection{Discussion}
\label{subsec:discuss}

Many important features of experimental SERS results for
benzenethiol \cite{VanDuyne09,Carron,SERS-Roshi,VanDuyne05} are
reflected in the computed Raman spectra of the PhS--Ag$_n$
complexes. The molecular vibrational frequencies of the complexes
are shifted compared to that of isolated benzenethiol. The
frequency shifts of intense $a'$ modes of the complexes
PhS--Ag$_7$ (I) and (II), PhS--Ag$_9$, and PhS--Ag$_{10}$ are
compared to experimental SERS data in Tab.~\ref{diss:table1}. The
values and the signs of the computed shifts are in good agreement
with the results of measurements. The magnitudes of the frequency
shifts depend on the degree to which the C--S bond is involved in
the given vibration. For instance, the C--S stretching mode
$\omega_3^{\rm exp}=1094$~cm$^{-1}$ shows the largest shift on
average, which is an additional manifestation of the proximity
effect discussed in Sec.~\ref{subsec:offres}.
\begin{table}
\begin{center}
\caption{Frequency shifts of benzenethiol Raman-active vibrational
modes due to binding to the metal. The values of the vibrational
frequencies are taken for neat benzenethiol from
Ref.~\cite{VanDuyne09}. The shifts for the computed complexes,
calculated as $\Delta\omega_n = \omega_n(\mbox{PhS--Ag$_k$}) -
\omega_n(\mbox{PhSH})$, are compared with the shifts of Raman
lines, $\Delta\omega_n = \omega_n(\mbox{SERS}) -
\omega_n(\mbox{PhSH})$, from the experimental studies,
Refs.~\cite{VanDuyne09,Carron,SERS-Roshi}. The frequencies are
given in cm$^{-1}$. }\label{diss:table1}
\begin{tabular}{|l|c|c|c|c|c|c|c|}
\hline
Mode & $\omega_1$ & $\omega_2$ & $\omega_3$ & $\omega_4$ & $\omega_5$ & $\omega_6$ & $\omega_7$ \\
\hline
Freq. PhSH & 1004 & 1027 & 1094 & 1583 & 414 & 701 & 918 \\
\hline
PhS--Ag$_7$ (I)         & -7 & -9 & -31 & -33 & 1 &  4 & -8  \\
PhS--Ag$_7$ (II)        & -4 & -5 & -24 & -17 & 3 &  6 & -12 \\
PhS--Ag$_9$            & -1 & -2 & -18 & -8  & 6 & -4 & -5  \\
PhS--Ag$_{10}$         & -1 & -3 & -18 & -9  & 5 & -4 & -11 \\
\hline
Ref.~\cite{VanDuyne09} & -1 &  0 & -18 & -7  & 8 & -6 &  -  \\
Ref.~\cite{Carron}     &  0 &  1 & -17 & -9  & 6 & -3 &  -  \\
Ref.~\cite{SERS-Roshi} & -1 & -2 & -21 & -11 & 5 & -8 &  -  \\
\hline
\end{tabular}
\end{center}
\end{table}

Comparisons of the computed Raman cross sections of the
PhS--Ag$_n$ complexes to experimental SERS data should take into
account that the chemisorbed molecules are no longer freely
rotating. This has the effect that the orientational averaging of
the Raman tensor depends on the shape of the metal substrate and
the relative orientation of the adsorbed molecules. As a model, we
consider benzenethiol molecules adsorbed on a spherical
nanoparticle. The computed PhS--Ag$_n$ structures are oriented
such that in the complexes with on-top binding, $n = 6,7$(I), the
Ag--S bond is orthogonal to the particle surface, and in the
complexes with the bridge binding motif, $n = 7$(II), $8-11$  the
plane formed by the two Ag--S bonds is orthogonal to the particle
surface. The Raman intensity of a vibrational mode with frequency
$\omega_n$ is~\cite{Creighton}
\begin{equation}
I(\omega_n,\omega_{\rm ex}) \propto (\omega_{\rm ex}-\omega_n)^4
|g(\omega_{\rm ex})|^2 |g(\omega_{\rm ex}-\omega_n)|^2
f(\alpha_n(\omega_{\rm ex}))\,, \label{Intens}
\end{equation}
where $\omega_{\rm ex}$ is the excitation frequency, and the
electromagnetic enhancement factor is
\begin{equation}
g(\omega)=\frac{\epsilon_{\rm p}(\omega)-\epsilon_{\rm
env}}{\epsilon_{\rm p}(\omega)+2\epsilon_{\rm env}}\,.
\label{emEF}
\end{equation}
$\epsilon_{\rm p}$ stands here for the dielectric constant of the
metal particle and includes the plasmon resonance, and
$\epsilon_{\rm env}$ is the dielectric constant of the
environment, which is considered as not strongly
frequency-dependent. $\alpha_n(\omega_{\rm ex})$ denotes the Raman
tensor of the mode $n$, and the orientational averaging function
is given by
\begin{equation}
 f(\alpha) =
4(\alpha_{11}^2+\alpha_{22}^2+16\alpha_{33}^2) + \alpha_{11}
\alpha_{22} + 4\alpha_{11} \alpha_{33} + 4 \alpha_{22} \alpha_{33}
+ 7(\alpha_{12}^2 + 4\alpha_{13}^2 +4\alpha_{23}^2)\,.
\label{falpha}
\end{equation}
The orientational averaging used here \cite{Creighton} is
different from the isotropic averaging that has been utilized in
the previous sections. Particularly convenient for comparison with
experiment are relative Raman scattering cross sections which are
defined as
\begin{equation}
\frac{I(\omega_n,\omega_{\rm ex})}{I(\omega_{\rm ref},\omega_{\rm
ex})} = \frac{g^2(\omega_{\rm ex}-\omega_n)}{g^2(\omega_{\rm
ex}-\omega_{\rm ref})} \cdot \frac{(\omega_{\rm ex}-\omega_n)^4
f(\alpha_n)}{(\omega_{\rm ex}-\omega_{\rm ref})^4 f(\alpha_{\rm
ref})}, \label{relI}
\end{equation}
where $\omega_{\rm ref}$ is a frequency of the reference
vibrational mode. The first multiplier on the right-hand side of
Eq.~\ref{relI} represents the relative electromagnetic enhancement
of the Raman scattering due to plasmon excitations in the metal
particle. The chemical bonding of a molecule to a surface modifies
the Raman tensor ${\alpha}$ and contributes to the second
multiplier of Eq.~\ref{relI}.
\begin{table}
\begin{center}
\caption{Relative intensities of Raman lines, defined as
$I(\omega_n)/I(\omega_2)$, $\omega_2 = 1027$~cm$^{-1}$, are given
for the PhS--Ag$_n$, $n=7,9,10$ complexes and for the experimental
SERS spectra from Refs.~\cite{VanDuyne09,Carron,SERS-Roshi}. The
relative intensities for the complexes are computed using
Eq.~\ref{relI} assuming that electromagnetic enhancement is
frequency independent, $g(\omega) = \mbox{const}$. The mode
$\omega_7 = 918$~cm$^{-1}$ is not observable in the SERS
experiments. The excitation wavelength is 785~nm. The vibrational
frequencies of neat benzenethiol \cite{VanDuyne09} are given in
cm$^{-1}$.}\label{diss:table2}
\begin{tabular}{|l|c|c|c|c|c|c|}
\hline
Mode & $\omega_1$ & $\omega_3$ & $\omega_4$ & $\omega_5$ & $\omega_6$ & $\omega_7$ \\
\hline
Freq. PhSH & 1004 & 1094 & 1583 & 414 & 701 & 918 \\
\hline
PhS--Ag$_7$ (I)     & 1.2 & 1.4 & 2.7 & 0.4 & 0.5 & 1.0 \\
PhS--Ag$_7$ (II)    & 1.2 & 1.4 & 0.5 & 0.1 & 0.2 & 1.6 \\
PhS--Ag$_9$        & 0.6 & 3.4 & 2.8 & 0.7 & 0.2 & 0.0 \\
PhS--Ag$_{10}$     & 0.6 & 5.1 & 4.5 & 1.1 & 0.3 & 0.0 \\
\hline
Ref.~\cite{VanDuyne09} & 1.3 & 1.4 & 0.6 & 0.6 & 0.3 & - \\
Ref.~\cite{Carron}     & 1.4 & 1.3 & 1.3 & 1.0 & 0.4 & - \\
Ref.~\cite{SERS-Roshi} & 1.1 & 1.0 & 3.5 & 2.7 & 0.4 & - \\
 \hline
\end{tabular}
\end{center}
\end{table}

The computed relative intensities of several Raman lines for the
complexes PhS--Ag$_7$ (I) and (II), PhS--Ag$_9$, and
PhS--Ag$_{10}$ are compared to the experimental data in
Tab.~\ref{diss:table2}. Experimental SERS cross sections depend
sensitively on the structure of the surface and the roughness
features of the SERS substrate and are thus strongly influenced by
the preparation technique, as illustrated by the differences
between the experimental data in Tab.~\ref{diss:table2}. Surface
enhanced Raman spectra result from the appropriate averaging over
the multitude of possible local bonding configurations. An
additional important effect is due to differences in plasmonic
resonance structures. Comparison of the experimental data with
finite-size cluster models PhS--Ag$_n$ shows that indeed many
experimental SERS features may be correlated with the models
considered here. For example, all strong Raman lines observed in
Ref.~\cite{VanDuyne09} belong to the $a'$ representation, similar
to the spectra of the PhS--Ag$_9$ and PhS--Ag$_{10}$ complexes.
However, the relative intensities of the $\omega_3^{\rm exp} =
1094$~cm$^{-1}$ and $\omega_4^{\rm exp} = 1583$~cm$^{-1}$ modes
are too high as compared to the experimental data. The discrepancy
stems from overestimation of the computed Raman intensities for
isolated benzenethiol and also from a frequency dependence of
electromagnetic enhancement, which is not accounted for in the
model. For instance, in Ref.~\cite{VanDuyne09} the excitation
frequency is optimized to enhance modes about
1000~cm$^{-1}$.~\cite{note1} Thus, the electromagnetic enhancement
of the $\omega_3$ and $\omega_4$ modes is weaker as compared to
the 1004~cm$^{-1}$ mode. This trend should reduce the discrepancy
between the experimental data of Ref.~\cite{VanDuyne09} and the
computed spectra of the PhS--Ag$_9$ and PhS--Ag$_{10}$ complexes.
On the other hand, Refs.~\cite{Carron} and \cite{SERS-Roshi} show
some enhancement of the vibrational modes at 470~cm$^{-1}$ and
960~cm$^{-1}$, respectively, which belong to the $a''$
representation. A possible explanation is that molecular
orientations on the metal surface have irregularities and also
contain binding situations similar to PhS--Ag$_7$ (I) and (II),
which show considerable enhancement for $a''$ modes. For instance,
the relative intensity of the 960~cm$^{-1}$ mode in PhS--Ag$_7$
(II) complex is 0.8.

To get an additional verification of our model, relative
enhancement factors for the 1094~cm$^{-1}$ and 1583~cm$^{-1}$
vibrational modes of the PhS--Ag$_7$ (I) and (II), PhS--Ag$_9$,
and PhS--Ag$_{10}$
complexes have been compared to the enhancement factors measured
in Ref.~\cite{VanDuyne05}. A relative enhancement factor is
defined for two vibrational modes $\omega_n$ and $\omega_{\rm
ref}$ with corresponding excitation frequencies $\omega_{\rm ex}$
and $\omega_{\rm ex'}$ as
\begin{equation}
\frac{{\rm EF}(\omega_n,\omega_{\rm ex})}{{\rm EF}(\omega_{\rm
ref},\omega_{\rm ex'})} = \frac{g^4(\omega_{\rm
ex})}{g^4(\omega_{\rm ex'})}\left(\frac{f(\alpha_n)}{f_{\rm
iso}(\alpha_n)}\right)/\left(\frac{f(\alpha_{\rm ref})}{f_{\rm
iso}(\alpha_{\rm ref})}\right), \label{relEF}
\end{equation}
where $f(\alpha)$ is calculated using Eq.~\ref{falpha}, $f_{\rm
iso}(\alpha)$ is a similar expression for an isotropic
case~\cite{Long}, and the excitation frequencies have been chosen
to get a maximal enhancement of corresponding vibrational
modes.~\cite{note1} Near a plasmonic resonance the first
multiplier on the right hand side of Eq.~\ref{relEF} can be
approximated by the ratio of the extinction coefficients of the
plasmonic structure. The computed relative enhancement factors for
the $\omega_3$ and $\omega_4$ modes of PhS--Ag$_n$, $n = 9,10$,
complex given in Tab.~\ref{diss:table3} agree well with the
experimental data from Ref.~\cite{VanDuyne05}.
\begin{table}
\begin{center}
\caption{Relative enhancement factors of molecular vibrational
modes computed for the PhS--Ag$_7$ (I) and (II), PhS--Ag$_9$ and
PhS--Ag$_{10}$ complexes using Eq.~\ref{relEF} as compared with
the experimental data from Refs.~\cite{VanDuyne05}. The reference
vibrational mode is $\omega_1^{\rm exp}(\mbox{SERS}) =
1009$~cm$^{-1}$. The corresponding excitation wavelength $\lambda
_{\rm ex}$ is in nm. To account for frequency dependence of
electromagnetic enhancement the computed values were multiplied by
$g^4(\omega_{\rm ex})/g^4(\omega_{\rm ex'})$ estimated from the
experiments. The vibrational frequencies in cm$^{-1}$ are for
benzenethiol on an Ag nanoparticle array from
Ref.~\cite{VanDuyne05}.}\label{diss:table3}
\begin{tabular}{|l|c|c|c|c|c|}
\hline
Mode & $\omega_1$ & $\omega_2$ & $\omega_3$ & $\omega_4$ \\
\hline
Freq. SERS                                   & 1009 & 1027 & 1081 & 1575 \\
$\lambda _{\rm ex}$                          & 705  & 703 & 700  & 692  \\
$g^4(\omega_{\rm ex})/g^4(\omega_{\rm ex'})$ & 1.0  & 1.0 & 0.9  & 0.6  \\
\hline
PhS--Ag$_7$ (I)  & 1.0 & 0.8 & 1.0 & 1.4  \\
PhS--Ag$_7$ (II) & 1.0 & 0.6 & 1.0 & 0.1  \\
PhS--Ag$_9$     & 1.0 & 1.7 & 5.3 & 2.9  \\
PhS--Ag$_{10}$  & 1.0 & 1.7 & 7.8 & 4.6 \\
\hline
Ref.~\cite{VanDuyne05} & 1 & - & 10 & 7 \\
 \hline
\end{tabular}
\end{center}
\end{table}

Resonant structure of Raman cross sections discussed in
Sec.~\ref{subsec:modes} is hardly observable in experiments of
Refs.~\cite{VanDuyne09,Carron,SERS-Roshi,VanDuyne05}, where Raman
scattering is measured for an inhomogeneous ensemble of molecules
with their local environment. However, it can be observed in
single molecule experiments. For example, the fine structure of
single-molecule REPs measured in Ref.~\cite{VD_sm} may be
interpreted as a result of resonances with mixed metal--molecular
states.

\section{Conclusions}
\label{sec:conclusions}

We have analyzed the Raman response of benzenethiol chemisorbed on
small silver clusters and find overall enhancement of the response
of the order of $10^3$ as compared to isolated benznethiol. Mixed
metal--molecular electronic states are formed due to chemical
bonding. Raman response from the metal--molecular complexes
exhibits resonance-type features even in the range far below the
onset of molecular electronic excitations.  We identify several
mechanisms contributing to enhancement of Raman scattering in both
off-resonant and resonant regimes. These include the relative
orientation of the benzene ring with respect to the cluster, the
local symmetry of the benzene ring, and the proximity of the
particular vibrational mode to the binding site. Additional
enhancement of Raman response by several orders of magnitude
arises from resonant excitations localized around the binding
sites, whereas response from other excitations in this range is
significantly weaker. Specific properties of Raman response are
strongly dependent on the local atomic environment of adsorbate.
However, important features in SERS of benzenethiol on silver
surfaces such as vibrational frequency shifts and relative
enhancement factors may be explained using finite-size cluster
models. Since the atomic structure of SERS substrates cannot
currently be controlled in experiments the described effects
should be averaged over a set of possible binding patterns.
Nevertheless, our results might provide a guidance for
optimization of SERS-active structures and are useful for
interpretation of single molecule SERS measurements.

\section{Acknowledgments}
We thank Roshan Aggarwal for providing the experimental data and
the text of the manuscript before its publication. We also much
appreciate Eric Diebold and Mohamad Banaee for useful discussions.
This work was supported by the Defense Advanced Research Project
Agency under Contract No FA9550-08-1-0285. R. O.-A. was supported
by CONACyT and Fundaci\'on M\'exico en Harvard A.C. We thank the
High Performance Technical Computing Center at the Faculty of Arts
and Sciences of Harvard University and the National Nanotechnology
Infrastructure Network Computation project for invaluable support.

\clearpage

\section{SUPPLEMENTARY INFORMATION}

\subsection{Electronic Excitation Spectra}

\begin{figure}[c]
  \begin{center}
  \centering
   \includegraphics[width=0.6\linewidth, clip=true]{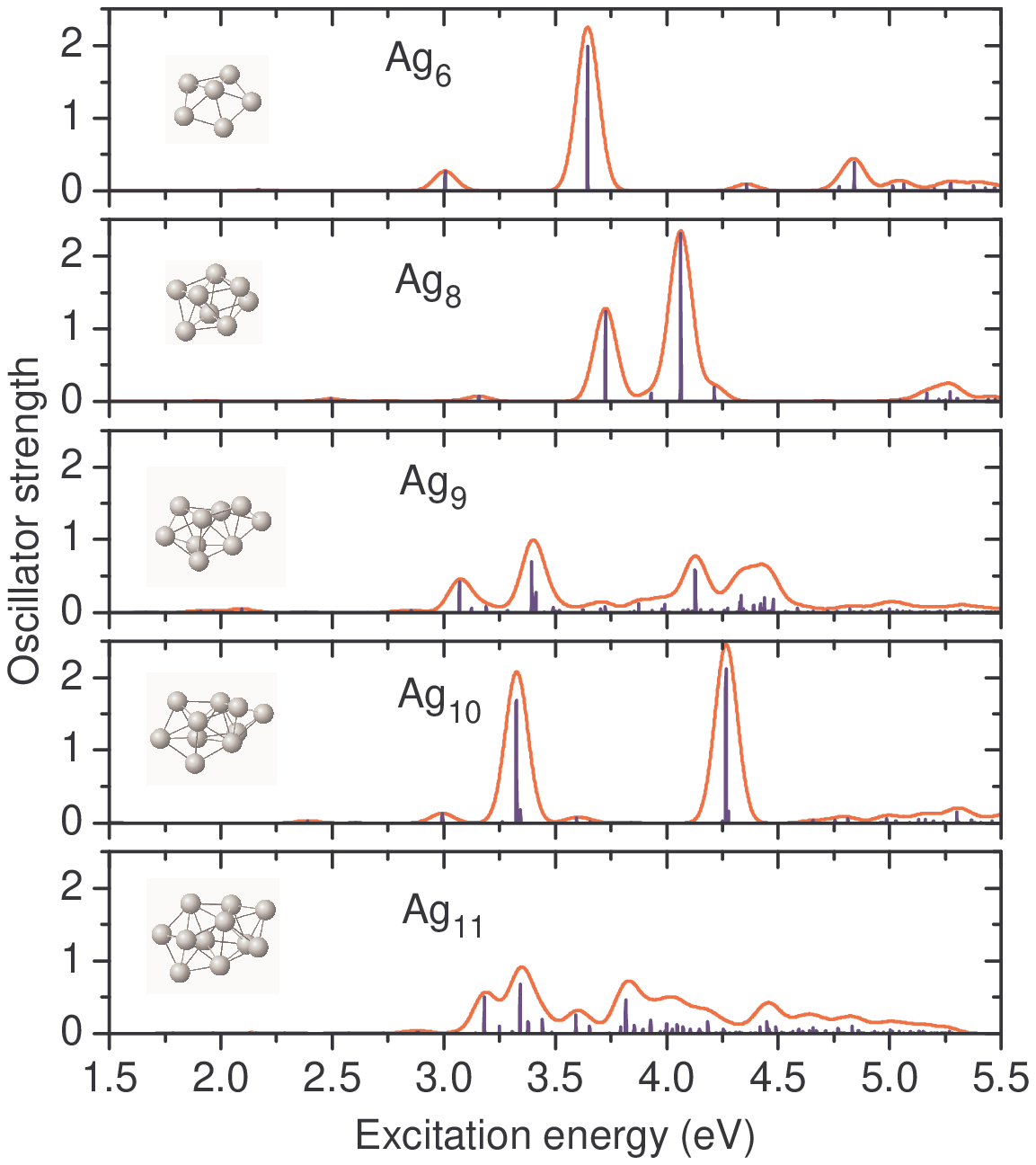}
  \caption{Absorption spectra of the computed Ag$_n$ clusters, $n=6,8-11$.}
    \label{fig1s}
  \end{center}
\end{figure}

\begin{figure}[c]
  \begin{center}
  \centering
   \includegraphics[width=0.6\linewidth, clip=true]{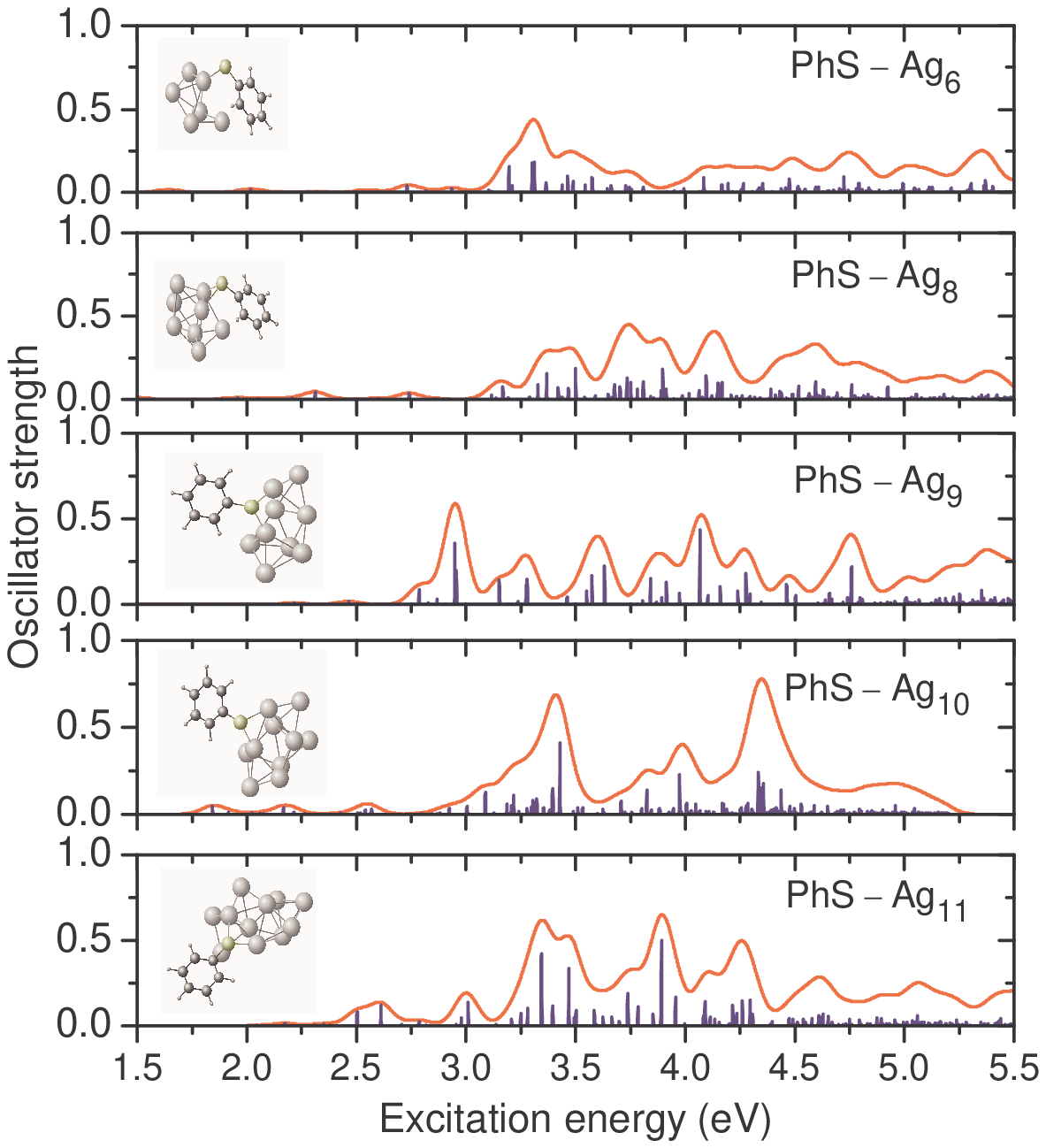}
  \caption{Absorption spectra of the computed PhS--Ag$_n$ complexes, $n=6,8-11$.}
    \label{fig2s}
  \end{center}
\end{figure}

The absorption spectra of the simulated Ag$_n$ clusters and
PhS--Ag$_n$ complexes, $n=6,8-11$ are shown in Figs.~\ref{fig1s}
and \ref{fig2s}. For $n=6$, the structure of the silver cluster in
the PhS--Ag$_n$ complex differs from the one of the bare silver
cluster because of the cluster reconstruction. This is discussed
in Sec.~3.1 of the manuscript in more details.

\subsection{Off-resonance Raman Spectra}

The relative intensities of the computed Raman lines of isolated
benzenethiol as compared to the relative Raman intensities
measured for neat benzenethiol \cite{VanDuyne09, Carron,
SERS-Roshi} are given in Tab.~\ref{tb1:SM}. The intensities are in
good agreement except of the $\omega_1^{\rm exp} = 1004$~cm$^{-1}$
vibrational mode. Changing the exchange functional in DFT
simulations results in minor improvements only.

\begin{figure}[c]
  \begin{center}
  \centering
   \includegraphics[width=0.6\linewidth, clip=true]{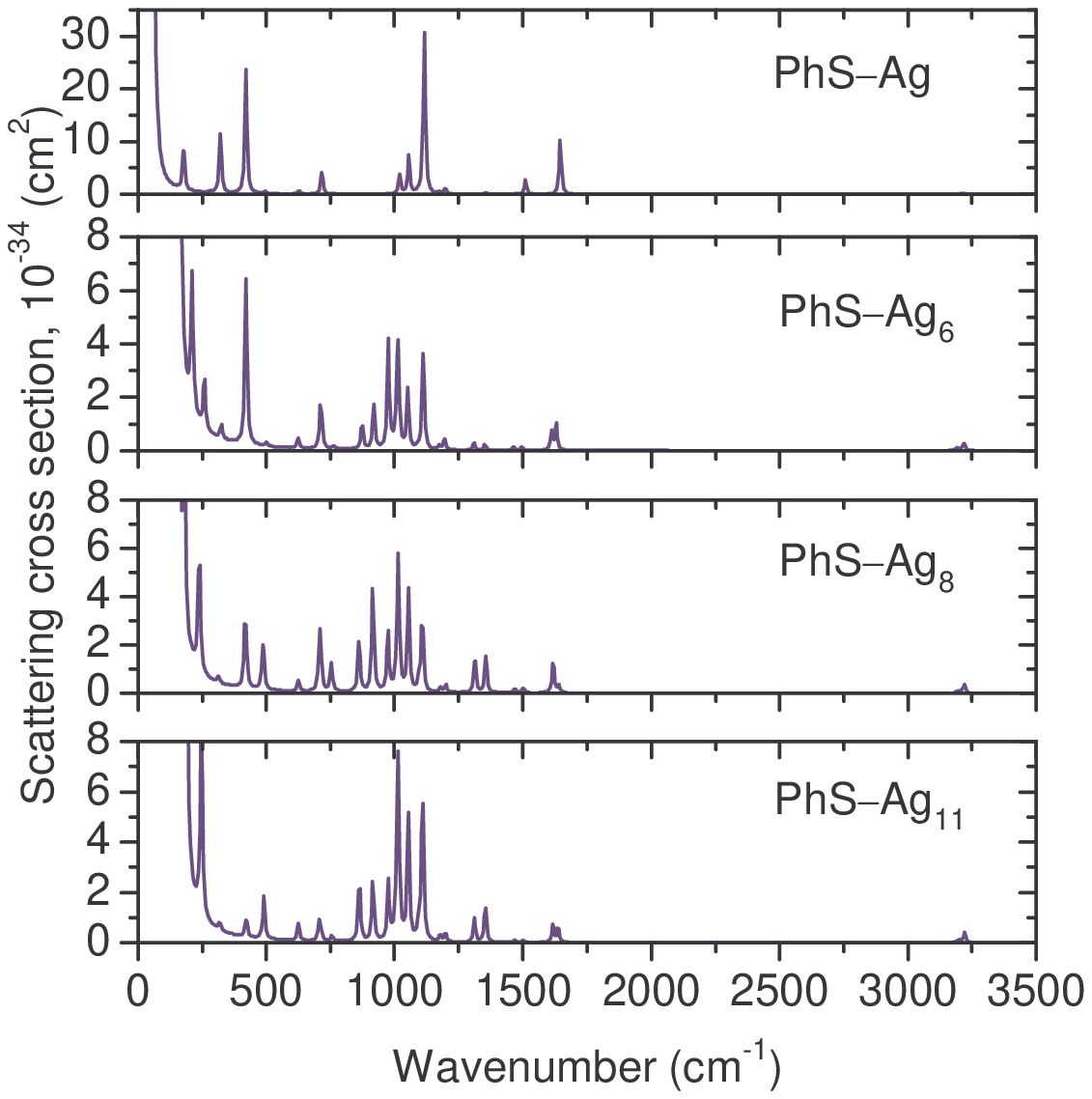}
  \caption{Off-resonance Raman spectra of the computed PhS--Ag$_n$
  complexes, $n=1,6,8,11$. The excitation energy is 0.62~eV, which corresponds to
  2000~nm wavelength.}
    \label{fig3s}
  \end{center}
\end{figure}

The Raman spectra of PhS--Ag$_n$ complexes, $n=1,6,8,11$ computed
at the excitation energy $0.62$~eV (2000~nm excitation wavelength)
are shown in Fig.~\ref{fig3s}. The excitation energy used is below
the lowest electronic transition in the structures. The complex
PhS--Ag shows a strong enhancement of totally symmetric
vibrational modes, while in the other complexes, the non-bonding
interaction between the aromatic ring and the cluster results in a
suppression of the integrated Raman scattering cross section and a
redistribution of intensities between the totally symmetric and
non-totally symmetric modes. See Sec.~3.2 of the manuscript.

\subsection{Raman Excitation Profiles}

\begin{figure}[c]
  \begin{center}
  \centering
   \includegraphics[width=0.6\linewidth,clip=true]{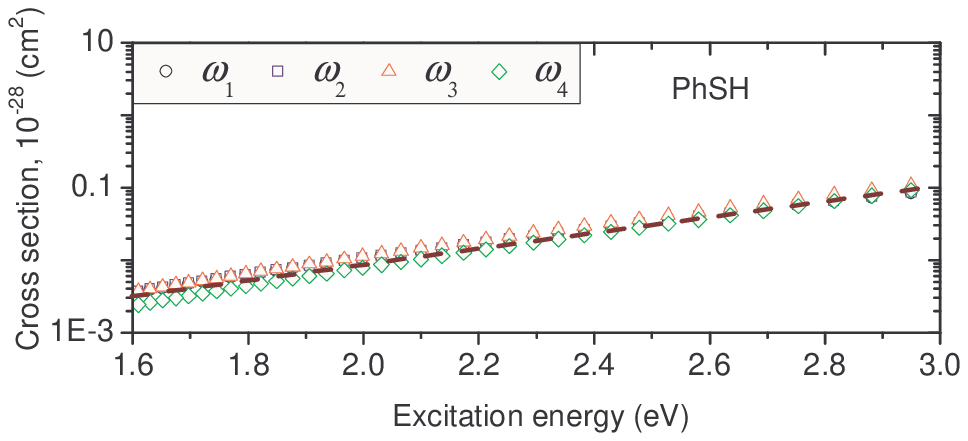}
  \caption{Raman excitation profile for the vibrational modes $\omega_1 =
1019$~cm$^{-1}$, $\omega_2 = 1059$~cm$^{-1}$, $\omega_3 =
1136$~cm$^{-1}$, and $\omega_4 = 1656$~cm$^{-1}$ in isolated
benzenethiol.}
    \label{fig4s}
  \end{center}
\end{figure}

\begin{figure}[c]
  \begin{center}
  \centering
   \includegraphics[width=0.6\linewidth,clip=true]{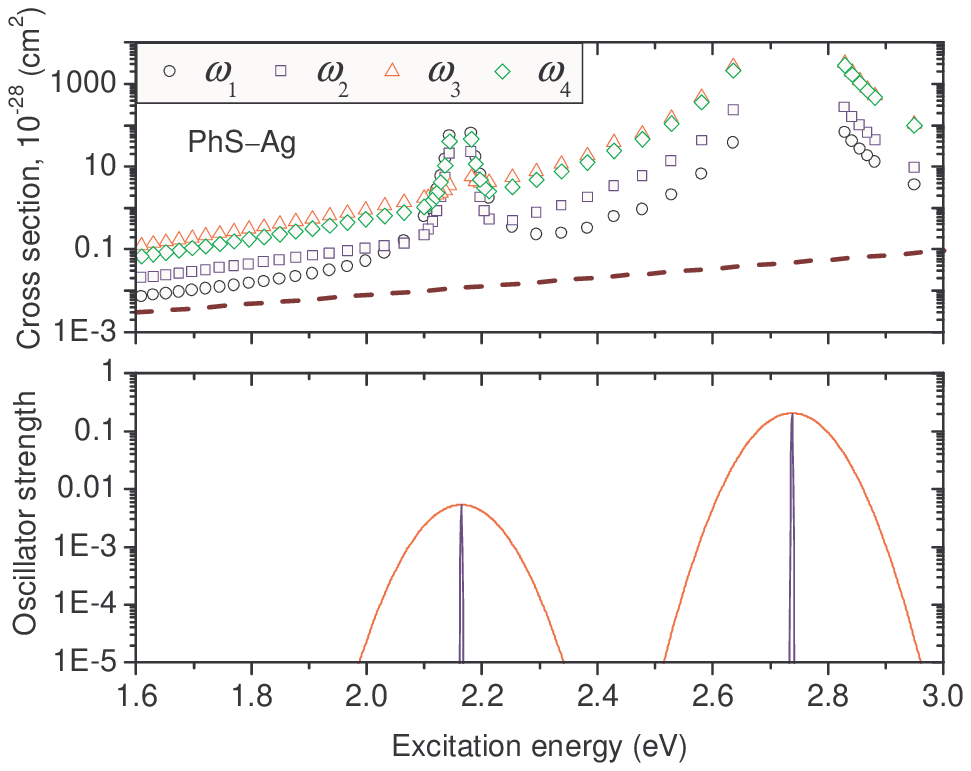}
  \caption{Raman excitation profile for the vibrational modes $\omega_1 =
1019$~cm$^{-1}$, $\omega_2 = 1059$~cm$^{-1}$, $\omega_3 =
1136$~cm$^{-1}$, and $\omega_4 = 1656$~cm$^{-1}$ in the PhS--Ag
complex.}
    \label{fig5s}
  \end{center}
\end{figure}

\begin{figure}[c]
  \begin{center}
  \centering
   \includegraphics[width=0.6\linewidth,clip=true]{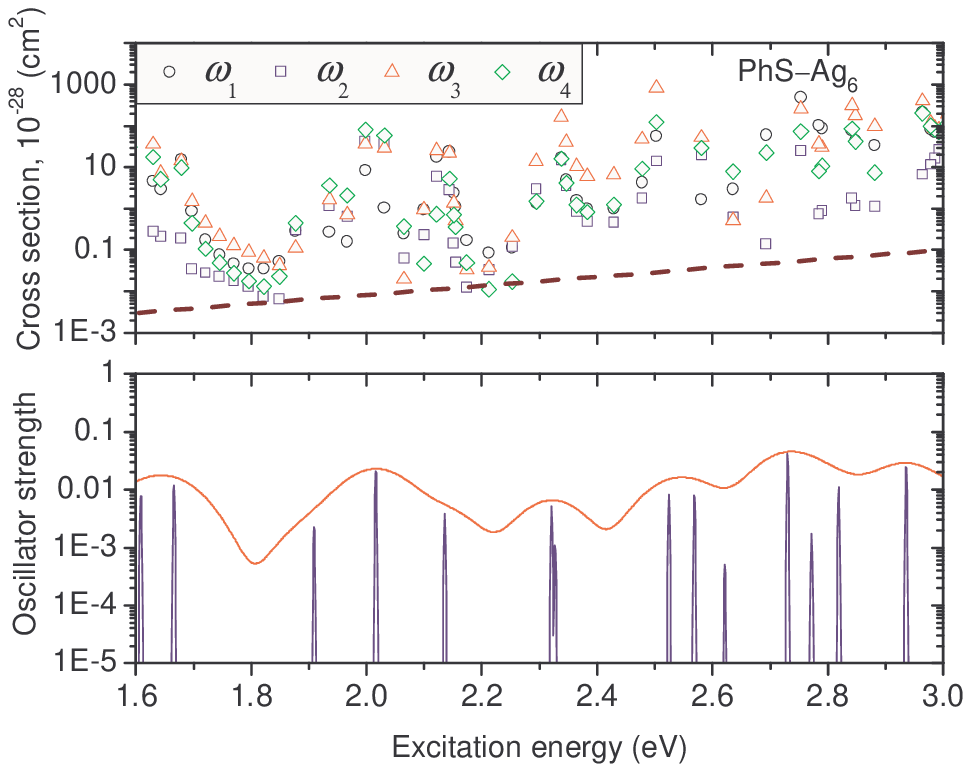}
  \caption{Raman excitation profile for the vibrational modes $\omega_1 =
1019$~cm$^{-1}$, $\omega_2 = 1059$~cm$^{-1}$, $\omega_3 =
1136$~cm$^{-1}$, and $\omega_4 = 1656$~cm$^{-1}$ in the
PhS--Ag$_6$ complex.}
    \label{fig6s}
  \end{center}
\end{figure}

\begin{figure}[c]
  \begin{center}
  \centering
   \includegraphics[width=0.6\linewidth,clip=true]{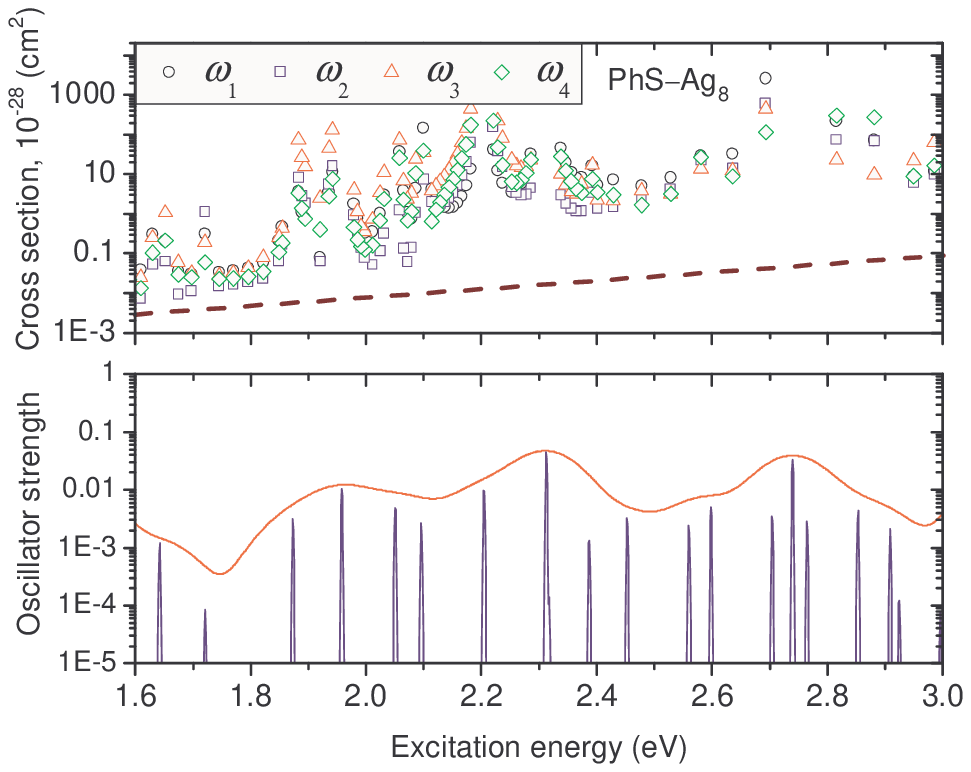}
  \caption{Raman excitation profile for the vibrational modes $\omega_1 =
1019$~cm$^{-1}$, $\omega_2 = 1059$~cm$^{-1}$, $\omega_3 =
1136$~cm$^{-1}$, and $\omega_4 = 1656$~cm$^{-1}$ in the
PhS--Ag$_8$ complex.}
    \label{fig7s}
  \end{center}
\end{figure}

\begin{figure}[c]
  \begin{center}
  \centering
   \includegraphics[width=0.6\linewidth,clip=true]{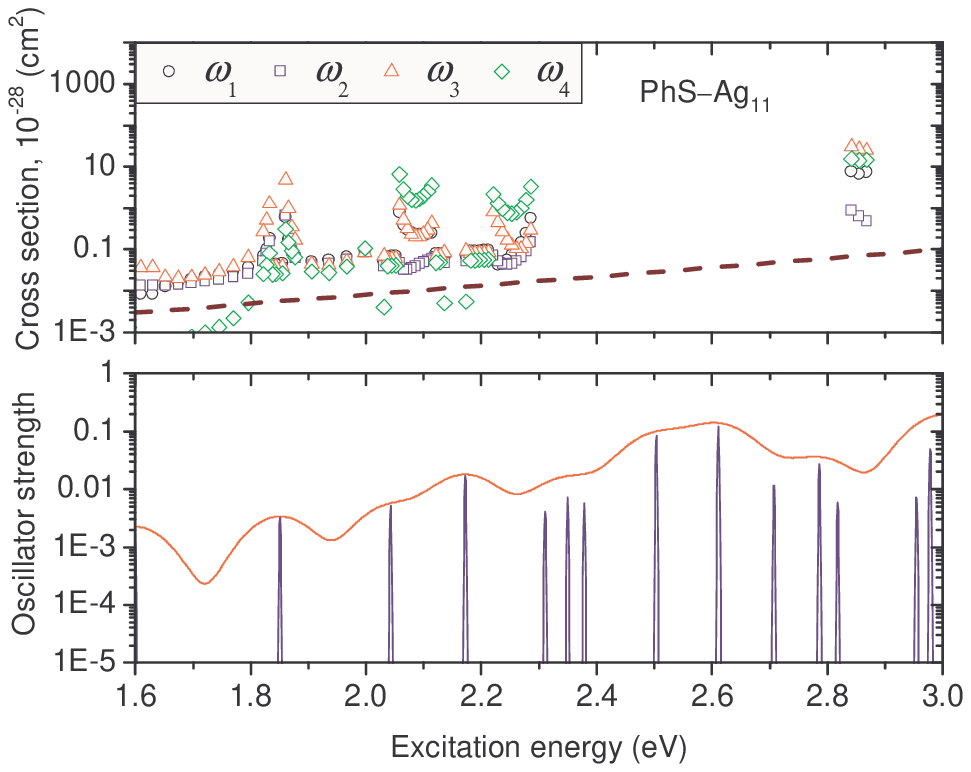}
  \caption{Raman excitation profile for the vibrational modes $\omega_1 =
1019$~cm$^{-1}$, $\omega_2 = 1059$~cm$^{-1}$, $\omega_3 =
1136$~cm$^{-1}$, and $\omega_4 = 1656$~cm$^{-1}$ in the
PhS--Ag$_{11}$ complex.}
    \label{fig8s}
  \end{center}
\end{figure}

The Raman excitation profiles (REPs) and low-frequency electron
excitation spectra for PhSH and PhS--Ag$_n$ complexes,
$n=1,6,8,11$ for the computed vibrational modes, $\omega_1 =
1019$~cm$^{-1}$, $\omega_2 = 1059$~cm$^{-1}$, $\omega_3 =
1136$~cm$^{-1}$, and $\omega_4 = 1656$~cm$^{-1}$, are shown in
Figs.~\ref{fig4s}-\ref{fig8s}. The range of excitation energies is
$1.6-3.0$~eV. In logarithmic scale the REPs for isolated PhSH,
Fig.~\ref{fig4s}, are almost linear in the excitation energy,
which is consistent with the semiclassical theory of Raman
scattering.\cite{Long} For the complexes PhS--Ag$_n$,
$n=1,6,8,11$, Figs.~\ref{fig5s}-\ref{fig8s},  a resonance-type
structure of REPs originates in the electronic excitations to
mixed metal-molecular states.

\subsection{Discussions}

The frequency shifts of four totally symmetric vibrational modes
$\omega_1=1019$~cm$^{-1}$, $\omega_2=1059$~cm$^{-1}$,
$\omega_3=1136$~cm$^{-1}$, and $\omega_4=1656$~cm$^{-1}$ in
PhS--Ag$_n$, $n= 1,6-11$ complexes are collected in
Tab.~\ref{tb2:SM}. On average, the values of the shifts are larger
for the complexes where total Raman scattering is quenched,
PhS--Ag$_n$, $n = 6,7,8,11$.

\newpage

\begin{table}
\begin{center}
\caption{The relative intensities of the computed Raman lines for
isolated benzenethiol as compared to the experimental data for
neat benzenethiol,
Refs.~\onlinecite{VanDuyne09,Carron,SERS-Roshi}. The relative
intensities are defined with respect to the 1027 cm$^{-1}$
vibrational mode. The excitation wavelength is 785~nm. The
frequencies are in
cm$^{-1}$.}\label{tb1:SM}
\begin{tabular}{|l|c|c|c|c|c|c|} \hline
Mode & $\omega_1$ & $\omega_3$ & $\omega_4$ & $\omega_5$ & $\omega_6$ & $\omega_7$ \\
\hline
Freq. PhSH & 1004 & 1094 & 1583 & 414 & 701 & 918 \\
\hline
PhSH$^{\rm comp}$  & 1.0 & 1.0 & 0.9 & 0.5 & 0.3 & 0.4 \\
PhSH$^{\rm exp}$ Ref.~\onlinecite{VanDuyne09} & 3.0 & 0.5 & 0.3 & 0.7 & 0.7 & 0.3 \\
PhSH$^{\rm exp}$ Ref.~\onlinecite{Carron} & 4.5 & 0.4 & 0.4 & 0.6 & 0.7 & 0.2 \\
PhSH$^{\rm exp}$ Ref.~\onlinecite{SERS-Roshi} & 2.9 & 0.6 & 1.0 & 1.0 & 0.5 & 0.4 \\
 \hline
\end{tabular}
\end{center}
\end{table}
\begin{table}
\begin{center}
\caption{The relative shifts of four totally symmetric vibrational
modes in PhS--Ag$_n$, $n = 1, 6-11$, complexes as compared to the
unperturbed frequencies in isolated benzenethiol, $\Delta\omega_n
= \omega_n(\mbox{PhS--Ag$_k$}) - \omega_n(\mbox{PhSH})$. The
computed frequencies for isolated benzenethiol are $\omega_1 =
1019$~cm$^{-1}$, $\omega_2 = 1059$~cm$^{-1}$, $\omega_3 =
1136$~cm$^{-1}$, and $\omega_4 = 1656$~cm$^{-1}$. The values of
the shifts are given in cm$^{-1}$.}\label{tb2:SM}
\begin{tabular}{|l|c|c|c|c|}
\hline
Complex &$\Delta\omega_1$ &$\Delta\omega_2$ &$\Delta\omega_3$ &$\Delta\omega_4$ \\
  \hline
 PhS--Ag$_1$     &-0&-4&-21&-10\\
 PhS--Ag$_6$     &-7&-8&-26&-28\\
 PhS--Ag$_7$(I)  &-7&-9&-31&-33\\
 PhS--Ag$_7$(II) &-4&-5&-24&-17\\
 PhS--Ag$_8$     &-5&-5&-29&-16\\
 PhS--Ag$_9$     &-1&-2&-18&-8\\
 PhS--Ag$_{10}$  &-1&-3&-18&-9\\
 PhS--Ag$_{11}$  &-6&-6&-28&-19\\
 \hline
\end{tabular}
\end{center}
\end{table}

\clearpage

\clearpage

\clearpage

\clearpage

\clearpage

\clearpage

\clearpage

\begin{thebibliography}{10}
\bibitem{note_err} The basis set superposition error of a pyridine
  molecule attached to an Ag$_{20}$ to be ca. 0.7~kcal/mol
  \cite{Schatz_JACS}.
\bibitem{note1} The enhancement factor for a given mode $\omega_n$ is maximal if
$\omega_{\rm ex}-\omega_{\rm pl} \approx \omega_n/2$, where
$\omega_{\rm pl}$ is the frequency of a plasmonic resonance.
\cite{VanDuyne05}  At this condition $g(\omega_{\rm ex}-\omega_n)
= g(\omega_{\rm ex})$.
\end{thebibliography}
\end{document}